\documentclass[11pt,letterpaper,USenglish,thm-restate]{article}

\usepackage{subcaption}

\usepackage{authblk}

\usepackage[dvipsnames]{xcolor}
\usepackage[margin=1in]{geometry}
\usepackage{setspace}
\usepackage{dsfont}
\usepackage{mathrsfs}
\usepackage{amssymb}
\usepackage{amsmath}
\usepackage{amsfonts}
\usepackage{amsthm}
\usepackage{graphicx}
\usepackage{mathtools}
\usepackage{verbatim}
\usepackage{cprotect}
\usepackage{algpseudocode}
\usepackage{pgfplots}
\usepackage[utf8]{inputenc}
\usepackage{listings}
\usepackage{algorithm}
\usepackage{algpseudocode}
\usepackage{xcolor}
\usepackage{hyperref}
\usepackage{enumitem}
\usepackage[capitalize]{cleveref}
\usepackage{array}
\usepackage{booktabs}
\usepackage{mathrsfs}

\usepackage{graphicx,wrapfig,lipsum}

\usepackage{thm-restate}

\usepackage{cite}

\newtheorem{definition}{Definition}
\newtheorem{theorem}{Theorem}
\newtheorem{lemma}{Lemma}

\newcommand{\vote}{\text{vote}}
\newcommand{\accuse}{\text{accuse}}
\newcommand{\bitc}{B}
\newcommand{\roundc}{R}
\newcommand{\vin}{v_{in}}
\newcommand{\gba}{\textsc{GBA}}
\newcommand{\wba}{\textsc{WBA}}
\newcommand{\ba}{\textsc{BA}}
\newcommand{\mvba}{\textsc{MVBA}}
\newcommand{\proofcrea}{\textsc{PC}}
\newcommand{\valdis}{\textsc{VD}}
\newcommand{\pval}{\textsc{PVA}}
\newcommand{\fullval}{\textsc{VA}}

\newtheorem{observation}{Observation}
\crefname{observation}{Observation}{observations}
\input{macros}

\title{Optimal Subcubic Byzantine Agreement with Plain PKI}
\author[]{Marc Dufay}
\author[]{Roger Wattenhofer}

\affil[]{ETH Z{\"u}rich\\ \{mdufay, wattenhofer\}@ethz.ch}

\date{\vspace{-5ex}}

\begin{document}

\maketitle
\begin{abstract}
We consider Byzantine Agreement (BA), where $n$ parties run a distributed protocol to agree on a common value, despite $t$ parties being byzantine. In the synchronous setting, a choice must be made between the weaker unauthenticated setting, which only allows $t < n/3$ byzantine parties, and the stronger authenticated setting that allows up to $t < n/2$ byzantine parties. For this latter setting, BA can be solved with optimal resiliency and optimal bit complexity. However, so far, all optimal solutions rely on the use of threshold signatures. This shows a gap compared to regular signatures, where only protocols with cubic complexity are known for optimal resiliency $t < n/2$.

We wonder whether this cubic complexity is required without the use of threshold signatures and show that this is not the case. More specifically, given $\kappa$ an upper bound on the size of a hash or signature, we describe a deterministic protocol solving BA with bit complexity $\mathcal{O}(\kappa^{(\sqrt{13} - 1)/2} n^{(\sqrt{13} + 1)/2}) \approx \mathcal{O}(\kappa^{1.303}n^{2.303})$ while having optimal resiliency $t<n/2$ and round complexity $\mathcal{O}(n)$.
    
\end{abstract}

\newpage
\tableofcontents
\section{Introduction}

Byzantine Agreement is a fundamental problem in distributed computing, where a set of parties must agree on a single value even in the presence of some malicious parties. In its simplest form, we are given $n$ parties, of which at most $t$ can be byzantine. Without any kind of setup, this problem can only be solved if $t < n/3$ \cite{unauthlowerbound}. However, if we assume the presence of a public key infrastructure (PKI), and more specifically that parties can sign messages, then we can solve the problem with $t < n/2$.

Another important aspect of the problem is the complexity of the protocol. In 1982, Dolev and Reischuk \cite{PODC:DolRei82} gave a quadratic lower bound for communication complexity. In the authenticated setting, this lower bound was only somewhat recently matched by Momose and Ren \cite{momoseren}. However, when considering optimal resiliency $t < n/2$, their protocol relies on threshold signatures \cite{JC:BonLynSha04}, which are a stronger cryptographic primitive, allowing parties, in certain cases, to compress $n$ signatures into one. Even though threshold signatures are used in practice, they have multiple drawbacks:
\begin{itemize}
    \item Threshold signatures require their keys to be correlated. A consequence is that a trusted dealer setup is needed. This is a strong assumption and requiring a trusted third party, even if only for the setup, for a decentralized protocol is not ideal. Another possible approach is to use a distributed key generation (DKG) protocol, but this is a complex protocol in itself.
    \item Looking in the future, most common signature and threshold signature primitives are not quantum-safe. While regular signatures quantum-safe alternatives exist and have already been thoroughly studied \cite{TCHES:DKLLSS18,sphincsquantsignature}, quantum-safe threshold signatures \cite{EC:DKMMPS24,C:EspNioPre24} are less mature and have been less studied.
    \item In general, it is interesting to known the best we can do with the least amount of cryptographic assumptions. 
\end{itemize}
For these reasons, we consider the problem of solving Byzantine Agreement with optimal resiliency $t < n/2$ with only regular signatures. To the best of our knowledge, all existing protocols for this setting have cubic complexity. In this work, we show that this is not a requirement and give the following result:

\begin{restatable}{theorem}{byzantineagreement}
    There exists a deterministic byzantine agreement protocol with round complexity $\roundc_{\ba}(n) = \mathcal{O}(n)$ and bit complexity $\bitc_{\ba}(n) = \mathcal{O}(\kappa^{(\sqrt{13} - 1)/2}n^{(\sqrt{13} + 1)/2}) \approx \mathcal{O}(\kappa^{1.303}n^{2.303})$.
\end{restatable}

Our work builds on the framework laid by Momose and Ren \cite{momoseren}, and make use of numerous new techniques, including more refined recursive protocols, improved uses of dispersers and fault-tolerant expander graphs, as well as fair cake cutting theory to achieve a subcubic complexity.

\paragraph{AI Disclosure:} ChatGPT-6 Astra was used to improve and refine ideas in \cref{sec:parallel-agreement,subsec:group-merging}. ChatGPT-6 Astra was also used to provide an initial draft, which the authors based their proof of \cref{theo:ba-full-round,theo:ba-final-comp}. Apart of the uses mentioned above, LLMs were only used for grammar and spelling checks. 

\section{Related work}

\paragraph{Deterministic synchronous protocols} In the deterministic synchronous setting, the Dolev-Reischuk \cite{PODC:DolRei82}  lower bound shows that a quadratic amount of communication is necessary to solve byzantine agreement. Dolev and Strong \cite{dolevstrong} also show that $t+1$ rounds are necessary to solve deterministic byzantine agreement. In the unauthenticated setting, Berman, Garay and Perry \cite{bitoptimalunauth} were the first to give a protocol with optimal resiliency $t < n/3$ and optimal bit complexity $\mathcal{O}(n^2)$. It took until $2020$ for Momose and Ren \cite{momoseren} to match this result in the unauthenticated setting. More precisely, they give a protocol with optimal resiliency $t < n/2$ and quadratic bit complexity $\mathcal{O}(\kappa n^2)$ using threshold signatures. When considering only regular signatures, they provide a protocol with resiliency $t < n/(2+\varepsilon)$ and bit complexity $\mathcal{O}(\kappa \varepsilon^{-1} n^2)$, which results in a cubic complexity for $t < n/2$.

\paragraph{Randomized protocols} We note that all the previous lower bounds only apply to deterministic algorithms. In the randomized settings, more efficient protocols exist, concerning both round and bit complexity \cite{PODC:DufPan25}. In particular, even in the much weaker asynchronous setting, when properties are only required to hold with high probability, byzantine agreement can be solved with expected near-linear complexity \cite{PODC:CohKeiSpi20,TCC:BKLL20}.

\paragraph{Bypassing the lower bound} There have been multiple attempts to bypass the deterministic lower bounds by using additional assumptions. Some of them rely on using adaptive algorithms which depend on the actual number of byzantine parties \cite{fromfewtomany,PODC:CDGGKV24}. Others rely on amortizing the cost of byzantine agreement over multiple instances \cite{PODC:CDGGKV25,convexagreement}. However, all approaches with resiliency $t \geq n/3$ rely on threshold signatures. A notable exception is \cite{convexagreement} which can be used to solve $\Theta(n)$ concurrent instances of byzantine agreement with an amortized near-linear cost per instance while only using regular signatures. However, this is only achieved with resiliency $t < n/(2 + \varepsilon)$.
\section{Overview}

Our main contribution is a new weak agreement protocol using plain PKI with subcubic message complexity. This protocol consists of $2$ main parts:
\begin{itemize}
    \item \textbf{Proof Gathering}: In the first part, parties exchange signed messages between themselves to get a proof that either $0$ or $1$ can be voted. We note that because of the lack of threshold signatures, a proof must contains at least $t+1$ signatures so that byzantine parties cannot forge it. In our protocol it is the case: a proof is made of exactly $t+1$ threshold signatures and thus has size $\Theta(\kappa n)$. At the end of this part, we require that at least $n/10$ honest parties have a proof for either $0$ or $1$. Additionally, if every honest party has the same input, then every honest party will get a proof for this input by the end of the proof gathering.
    \item \textbf{Validated Dispersion}: From the first part, at least $n/10$ honest parties have a proof. Our goal in this part is to make every honest party hold the same proof, i.e that there is a specific $proof_{v}$ with $v \in \{0,1\}$ such that every honest party has received $proof_{v}$ by the end of this part.
\end{itemize}

After running consecutively these two parts, if a party has a proof for both $0$ and $1$, it will decide $\bot$. Otherwise, it will decide the value matching the proof it received (validated dispersion ensures at least one proof will be received). Although these two sections sound simple to explain, obtaining a subcubic protocol for both of them is the main difficulty of this paper and requires multiple complex ideas. An overview is given below:

\subsection*{Proof construction}

The goal of proof gathering is that, by the end of it, at least $n/10$ honest parties have a proof showing that either $0$ or $1$ can be decided. We first give an initial idea for the proof. A proof for $v \in \{0,1\}$ allows a party to confirm it can decide $v$ without any additional communication. In particular, if all honest parties have input $v$, a proof for $1-v$ cannot exist. A common construction for such a proof is to have every party sign and send its input. If a party gets $t+1$ distinct signatures for $v \in \{0,1\}$, it shows that at least one honest party had input $v$ and thus that this value can be decided. 

Using threshold signatures, such a proof can be encoded using $\mathcal{O}(\kappa)$ bits. Here, because only plain signatures are available, such a proof will require $(t+1)\kappa = \mathcal{O}(\kappa n)$ bits. As such, a protocol should be careful when sending this proof to avoid having a cubic complexity. We present the different ideas used to achieve proof gathering with $\mathcal{\tilde{O}}( \kappa n^2)$ bit complexity and $\mathcal{O}(\log n)$ rounds complexity:

\paragraph{I know that I can't know:} As said before, we consider a first round where every party votes for its input and sends its (signed) vote to every other party. Given a party $v$, this can results in two outcomes for $v$:
\begin{itemize}
    \item $v$ gets enough votes for $0$ or $1$ to get a proof for either $0$ or $1$
    \item $v$ is not able to get a proof for either $0$ or $1$. However, in this case, we make the observation that $v$ still gained information from this step: it learned that all honest parties don't hold the same value (otherwise it would have received $n-t \geq t+1$ votes for this value), and therefore both $0$ and $1$ are "fine" to decide. To symbolize this, it sets its input to $\bot$. A downside of this approach is that $v$ has no way to prove this observation to other parties. But we show that this simple remark is powerful enough to always get a proof afterward. We note that this approach was already used to obtain optimal adaptive byzantine agreement protocol using threshold signatures \cite{fromfewtomany,optimaladaptivemulti}.
\end{itemize}

Let $nb_{\bot}$ be the number of honest parties which set their input to $\bot$ (they know that both values are fine). We remark that if $nb_{\bot} \leq 4n/10$, then at least $n - t - nb_{\bot} \geq n/10$ honest parties have a proof and therefore this step is already done.

In the case where $nb_{\bot} > 4n/10$, we perform a second round of input (which can now be $0$, $1$ or $\bot$) signing and sending. We now observe that byzantine parties can only behave in two different ways.

\paragraph{If byzantine parties behave as honest:} We consider the first case, where multiple byzantine will send their vote to an honest party, as a way to appear as honest from this party's point of view. More specifically, we consider the case when any honest party receives at least $7n/10$ votes. In this case, we use a disperser to disperse the votes received. Every vote is forwarded to a constant (different for each vote) number of parties. This ensures that any set of $4n/10$ honest parties holds enough information together to produce a proof. Afterwards, we give and use the following fault-tolerant expander, which we believe to be of independent interest:

\begin{restatable}{theorem}{expander}\label{theo:fault-tol-exp}
    For any $\alpha\in ]0,1[$, $\beta > 1$, there exists $D > 0$ such that for any $n > 0$, there exists an explicit graph $G = (V,E)$ on $n$ vertices of bounded degree $D$. Additionally, for any $C \subset V$ such that $|C| \leq \alpha \cdot n$, there exists $S \subseteq V$ such  that $S \cap C = \emptyset$, $|S| \geq n - \beta \cdot |C|$, and the induced subgraph $G[S]$ is connected and has diameter $\mathcal{O}(\log n)$. 
\end{restatable}

We note that this is a generalization and stronger version of \cite{PODC:Upfal92}'s result, which had fixed $\alpha = 1/72$ and $\beta = 6$. Using this graph, we ensure that a subset of $4n/10$ honest parties will be able to communicate and all get a proof in $\mathcal{O}(\log n)$ rounds.

\paragraph{If byzantine parties behave as malicious:} If the previous case is not reached, it implies that all honest parties received less than $7n/10$ votes and thus each honest party can identify at least $3n/10$ byzantine parties. We note that in this case, votes only may not be enough to get a proof. Therefore, we combine votes with our next "trick": accusations. An honest party can only accuse another party if it is byzantine. Thus, given a signed accusation from $p_i$ to $p_j$, we know that either $p_i$ or $p_j$ is byzantine. This approach is inspired by the work done for repeated agreement \cite{PODC:CDGGKV25} and relies on the insight formalized by \cite{AC:HirMauPrz00}: instead of trying to determine whether $p_i$ or $p_j$ is byzantine, we can just discard the two parties. 

This is implemented by adding accusations into a proof. For example, with 
$n = 5$ and $t=2$, assume the proof consists of: $\{ \vote_{p_1}(0), \vote_{p_2}(\bot), \accuse(p_3, p_4)\}$, we know that either $p_3$ or $p_4$ is byzantine, so at most one party among $p_1$ and $p_2$ is byzantine and thus $0$ is a valid value. We note that for this to work, votes and accusations must be disjoint, and in particular, accusations must form a matching. 

\paragraph{From accusations to fair cake-cutting:} In this setting, each honest party wants to get "rid of" up to $3n/10$ non-$\bot$ votes, and each honest party can accuse at least$3n/10$ byzantine parties. We view this in another way: consider the set of $n$ parties to be a cake. Each party knows at least $3n/10$ byzantine parties which we identify as delicious slices. The goal becomes for a party to split the cake such that each relevant party is given a range which contains what it considers to be a delicious slice: this is fair cake-cutting theory. We adapt an existing algorithm to our setting and show that such an assignment can always be obtained by asking each party to cut a range of the cake into $2$ equal parts $\lceil \log n \rceil$ times. This approach has bit complexity $\mathcal{O}(n^2 \log^2 n)$ and round complexity $\mathcal{O}(\log n)$.

\subsection*{Validated dispersal}

Using proof construction, we can guarantee that at least $n/10$ honest parties hold a proof (possibly not for the same value). Each party running the proof dispersal algorithm will get a set of proofs in return. The main property of proof dispersal is that there exists a proof which is in every set returned by an honest party. This implies that every party will either get a proof for $0$ or every party will get a proof for $1$. Because a proof for $v$ can only exist if $v$ satisfies weak validity (i.e if all honest party don't have input $1-v$), we show that deciding $v$ if a party only receives proofs for $v$ or otherwise deciding $\bot$ satisfies weak agreement.

Validated dispersal in its definition does not consider proofs, it consider any $L$-bit input which satisfies some external validity predicate (where in our case, the validity predicate is that the value is a valid proof). We note that a really simple implementation of validated dispersal can be to have every party send its input (if it has one) to every other party, then return the set of inputs received. This works and can be done in one round, however, it has cubic complexity. 

\paragraph{Divide-And-Conquer}: The approach we took to improve the complexity in this paper is based on a divide-and-conquer approach (and later recursive calls). We divide the $n$ parties into $m := m(n)$ groups of similar size (where $m$ is decided later) and make two key remarks. The first remark is that by sending messages using a disperser pattern, we can make it so at most $n/(10m)$ honest parties do not have an input, with a bit cost of $\mathcal{O}(n^2 m)$. The second insight is that if one of the groups has a byzantine minority (which we are guaranteed at least one has), and all honest parties in this group agree on the same value, then, using standard erasure coding techniques, all parties can learn this value using only $\mathcal{O}(n^2)$ communication. Summing this over all $m$ groups, we get the same $\mathcal{O}(n^2 m)$ bit cost.

\paragraph{Validated Agreement:}
Using the previous idea, we can effectively reduce Validated dispersal to $m$ parallel calls to agreement on a valid value within a group of smaller size $n/m$. We note that this problem is not exactly regular Validated Byzantine Agreement, as some honest parties may not have an input. We call this problem Partial Validated Agreement. Nevertheless, we show that Partial Validated Agreement can be implemented using $\mathcal{O}(\log n)$ calls to regular Validated Agreement.

Our setting is now Validated Byzantine Agreement with a byzantine minority. An easy way to solve this would be to have every party send its input using Byzantine Broadcast (for example the Dolev-Strong protocol). Although this simple approach would be enough to achieve a subcubic complexity for the overall protocol, there exists better approaches. 

A first improvement would be to implement Byzantine Broadcast using our own Byzantine Agreement protocol, effectively making our protocol recursive. This is more efficient, but requires $n$ recursive calls in total and also has a high bit-complexity due to multiple inputs sharings.

\paragraph{Dynamic Programming approach}
When considering the previous Validated Agreement running with $n$ parties and input size $L$, it requires $\mathcal{O}(\kappa n)$ recursive binary agreement calls and its dependency on $L$ is $\mathcal{O}(n^2L)$. To reduce the dependency on $L$ to $\mathcal{O}(n L)$, we take a dynamic programming approach: for a set $S$ of parties, we keep track of a tuple $(d, H)$ where $d$ is the Merkel root of a valid value and $H$ is a set of parties that know the value associated with $H$. One key invariant we keep is that $S \ H$ has a byzantine majority (i.e at least half of its parties are byzantine). We show how to merge two disjoint sets $S$ and $S'$ where we have a tuple satisfying these properties to obtain a new tuple for $S \cup S'$ using only $|S'|$ byzantine agreement calls with an additional $\mathcal{O}(L |S'|)$ bit complexity. When merging sets in the correct order, this allows the protocol to only have an $\mathcal{O}(nL)$ dependency on $L$.

\paragraph{Batching}
Our dynamic programming approach relies on running multiple independent byzantine broadcast procedures. Each broadcast requires one recursive call to our byzantine agreement protocol, which is expensive. To work around this issue, we show how to use batching to improve the overall bit complexity. Let $b := b(n)$, we consider $b$ independent instances of byzantine broadcast from a set of broadcasters $S$. We rely on a committee-based broadcast protocol, based on the Dolev-Strong protocol, which is more efficient, but only guarantees agreement if at least one party in the committee is honest. Using $S$ as the committee, each party tries to broadcast its value using the committee-based protocol. Afterwards, a single multi-valued byzantine protocol (implemented on top of our binary BA protocol) is run to check the consistency of all $b$ instances at once. This decreases the number of recursive calls needed from $\mathcal{O}(\kappa n)$ to $\mathcal{O}(\kappa n/b)$.

\paragraph{Parallel execution}
A point we did not mention up to now is round complexity. Byzantine Agreement requires $\Omega(n)$ rounds and our goal is for our algorithm to be optimal in this regard.
When implementing the ideas presented above naively while optimizing $m$ and $b$ to minimize the bit complexity, the resulting algorithm would end up with a super-linear round complexity. To solve this, we make a more efficient use of the batching approach above, in particular by batching byzantine broadcasts from multiple independent merges together. This, together with running independent subprotocols in parallel whenever possible, allows the final round complexity to be linear.

\paragraph{Overall Complexity}

As said before, our algorithm is recursive: it recursively calls itself on smaller groups. When computing the bit complexity, we show that the per-party bit complexity $B_{BA}(n)$, while optimizing the parameters $m$ and $n$ is:

\begin{align*}
    B_{BA}(n) = 2B_{BA}(n/2) + \mathcal{\tilde O}(\kappa n) + \min_{1 \leq b \leq n/m < n} \mathcal{\tilde{O}} \left( n m +  \kappa n b^2 / m + \frac{\kappa n}{b^2 m} + \frac{\kappa n}{mb}B_{BA}(n/m) \right)
\end{align*}

We get the optimal value by choosing $b = m = \tilde \Theta((\kappa n)^{(\sqrt{13} - 3)/2})$ which leads to the algorithm per-party bit complexity $B_{BA}(n) =\mathcal{\tilde{O}}(n^{(\sqrt{13}-1)/2})$.

Regarding round complexity, the recurrence relation is:
\begin{align*}
    R_{BA}(n) = 2R_{BA}(n/2) + \mathcal{O}(\log n) + m^{1/3} R_{BA}(n/m)
\end{align*}
For our optimal choice of $m$, the resulting round complexity is the expected optimal $R_{BA}(n) = \mathcal{O}(n)$.
\section{Preliminaries}

We consider a set $\mathcal{P}$ of $n$ parties, out of which up to $t$ can be byzantine. We focus on fully deterministic protocols. The network is fully connected and channels are authenticated: any party can send a message to any other party, and the receiver will know who the sender is.

\paragraph{Model:} We consider a synchronous model, where parties communicate in rounds. In each round, parties can send messages to other parties, and all messages will be received by the beginning of the next round. 

\paragraph{Hashes:} We assume the existence of a cryptographic hash function that is collision-resistant: it is computationally infeasible to find two distinct inputs that hash to the same output. We assume that the output of the hash function is $\kappa$ bits long. Because byzantine parties should not be able to brute force a hash, we assume that $\kappa = \Omega(\log n)$. Another assumption made in this work is that $\kappa = \text{polylog}(n)$.

\paragraph{Signatures:} Each party has a public-private key pair, where the public key is known to all parties, and can sign messages with its private key. We assume that signatures are unforgeable: it is computationally infeasible for a party to forge a signature of another party. The signature size is $\kappa$ bits.

\paragraph{Erasure Coding:} Erasure coding allows a message to be split into multiple shares, such that a subset of the shares can be used to reconstruct the original message. We use classical Reed-Solomon codes \cite{reedsolomon}:
\begin{definition}
    An $L$-bit value can be split into $m$ shares, such that any subset of $m/2$ shares can be used to reconstruct the original value. Each share has size $\mathcal{O}(L/m + \log m)$ bits.
\end{definition}

\paragraph{Cryptographic Accumulator/Merkel Tree:} Our work relies on the use of cryptographic accumulators. Given a set $S$ of $m$ elements, a cryptographic accumulator returns an accumulator value $h$ for $S$ as well as a witness $w_s$ for each element $s \in S$ such that given the accumulator $h$, an element $s$ and its witness $w_s$, a party can verify that $s \in S$ without knowledge of $S$. When implemented using a Merkel Hash Tree \cite{merkeltree}, the accumulator has size $\kappa$ while each witness has size $\kappa \lceil \log n \rceil$.

We note that more efficient cryptographic accumulator implementations exist \cite{derler2015ca}, however they rely on stronger cryptographic primitives, and a more complex setup, which our work aims to avoid.

We always use erasure coding paired with cryptographic accumulators: When generating shares of a value, each share, along with its index, is considered as an element of a set, and the accumulator is computed on this set. This way, a party can verify that a share is valid using its witness.
\section{Definitions}

All the following protocols are assumed to be run on $n$ parties, out of which at most $t < n/2$ are byzantine.

\begin{definition}
    A protocol where each party is given an input $v_{in}$ and decides an output $v_{out}$ satisfies byzantine agreement if:
    \begin{itemize}
        \item Termination: All honest parties decide
        \item Validity: If all honest parties have the same input $v$, then only $v$ can be decided
        \item Agreement: If two honest parties decide respectively $v$ and $v'$, then $v = v'$.
    \end{itemize}
\end{definition}

Even though we give (and use) protocols for byzantine agreement where the input is an $L$-bit value, the target of this paper is binary agreement, where $\vin \in \{0,1\}$. 

\begin{definition}
    A protocol where each party is given an input $v_{in} \in \{0,1\}$ and decides an output $v_{out} \in \{0,1,\bot\}$ satisfies weak byzantine agreement if:
    \begin{itemize}
        \item Termination: All honest parties decide
        \item Validity: If all honest parties have the same input $v$, then only $v$ can be decided
        \item Weak Agreement: If two honest parties decide respectively $v \ne \bot$ and $v'$, then $v' = v$ or $v' = \bot$.
    \end{itemize}
\end{definition}

\begin{definition}
    We consider an external validity function \textsc{val} and a protocol where parties decide an output $v_{out}$ with the following properties:
    \begin{itemize}
        \item Termination: All honest parties decide
        \item Validity: If an honest party decides $v$, then $v$ satisfies the validity function
        \item Agreement: If two honest parties decide respectively $v$ and $v'$, then $v = v'$.
    \end{itemize}
    If every honest party is given an input $v_{in}$ satisfying the validity function, we say that the protocol satisfies validated agreement.

    If at least $n/10$ honest parties are given an input $v_{in}$ satisfying the validity function, we say that the protocol satisfies partial validated agreement.
\end{definition}

\paragraph{Byzantine minority/majority:} When considering a set of $Q$ parties, out of which $k$ are byzantine, we say that $Q$ has a byzantine minority if $k < |Q|/2$, and a byzantine majority otherwise (i.e when $k \geq |Q|/2$).

\paragraph{Proofs:} Our protocols rely on the creation and dissemination of "\textit{proofs}", which allows parties to verify whether or not they can decide a given value.

\begin{definition}\label{def:proof}
    Assuming all honest parties have an input in $\{0,1\}$, it is computationally infeasible for an adversary to compute a proof for $v \in \{0,1\}$ if all honest parties have input $1-v$.
\end{definition}

\paragraph{Byzantine-robust protocols:} Our protocols are designed to be able to solve byzantine agreement with resiliency $t < n/2$, which is optimal in regard to this problem, even with a trusted setup (if all honest parties have input $0$ and all byzantine parties act as honest with input $1$, BA cannot be solved if $t \geq n/2$). However, one of the key points of our protocol is that we partition the $n$ parties into smaller groups then run recursively our algorithm on them. This approach guarantees that at least one of the groups has an honest majority, but this may not be the case of the remaining ones. Although we do not expect anything from the output of protocols run in a byzantine majority, messages sent in this context by honest parties still count toward to the total bit complexity. To prevent any issue in this regard, we define the term of \textbf{byzantine-robust} protocols:

\begin{definition}
    Given an upper bound on the bit complexity per party, a protocol is byzantine-robust if this upper bound is satisfied regardless of the messages received. In particular, this applies even if all other parties are byzantine.
\end{definition}

A byzantine-robust protocol must be balanced, as such it makes more sense to look at the per-party bit complexity (i.e the number of bits sent by a party) rather than the total bit complexity. Given a protocol $\Pi$, we denote by $\bitc_\Pi(n)$ the per-party bit complexity of $\Pi$ when run on $n$ parties. We also denote by $\roundc_\Pi(n)$ the round complexity of $\Pi$ when run on $n$ parties.

Because all of our primitives and approaches have a bit cost increasing with the number of parties, we make simplifying assumptions that all our protocols complexities are non-decreasing with the number of parties. This allows us to simplify the analysis of our protocols.

\paragraph{Graphs:} Given a (possibly directed) graph $G = (V,E)$, the neighborhood of a vertex $u \in V$ is $N_G(u) = \{ v \in V, u \rightarrow v \in E \}$. Given a set $S \subseteq V$, we denote $N_G(S) = \bigcup_{u \in S} N_G(u)$.

Our work makes multiple uses of dispersers, which are defined below:
\begin{definition}
    A graph $G = (V,E)$ is a $(a, b)$-disperser if for any subset $S \subseteq V$ of size $|S| \geq a$, we have $|N_G(S)| \geq b$.

    A bipartite graph $G = (L \cup R, E)$ is a bipartite $(a, b)$-disperser if for any subset $S \subseteq L$ of size $|S| \geq a$, we have $|N_G(S)| \geq b$.
\end{definition}

A last point to remark is that many of our protocols are designed to run in parallel, but there is not enough information in the messages to know which instance to target. Moreover, our protocols, as they are written, are weak to replay attacks. We keep them this way for simplicity, but we show in \cref{sec:tag} how to add tags to messages and signatures to fix these issues, while keeping the same complexity.
\section{Weak byzantine Agreement}

Our Byzantine Agreement protocol uses the framework introduced by Momose and Ren \cite{momoseren}, who show how to implement Byzantine Agreement on top of Weak Agreement. Therefore, our focus and main contribution is implementing a weak agreement protocol using plain PKI. For completeness, protocols showing the reduction from byzantine agreement to weak agreement are provided in \cref{subsec:ba-protocol,subsec:gba-protocol} . Our protocol relies on two subprotocols: \textbf{proof creation} which makes sure at least $n/10$ honest parties are able to get ahold of a proof and \textbf{Validated dispersion} which makes sure there exists one proof received by every party. We provide formal definitions of these two subprotocols:

\begin{definition}
    In a proof creation protocol, each party has an input $v_{in} \in \{0,1\}$ and outputs a message $m$ such that the following properties are satisfied:
    \begin{itemize}
        \item Termination: All honest parties output.
        \item Certifiability: If an honest party outputs a message $m \ne \bot$, then $m$ is a proof for either $0$ or $1$, relevant to the inputs.
        \item Partial decision: At least $n/10$ honest parties do not output $\bot$.
    \end{itemize}
\end{definition}

\begin{definition}
    In a validated dispersal protocol, given some external validity function, each party is given an input which is either $\bot$ or an $L$-bit message satisfying the validity function. Additionally, at least $n/10$ honest parties do not have $\bot$ as input. Each party outputs a set $S$ of messages such that:
    \begin{itemize}
        \item Termination: All honest parties output.
        \item Validity: All messages output by an honest party satisfy the validity function.
        \item Coverage: The exists a valid $L$-bit message $m$ such that $m$ is contained in the set of every honest party's output.
    \end{itemize}
\end{definition}

Assuming access to these two subprotocols, we provide a simple implementation of weak agreement. The validity function used for validated dispersal is to check whether the message is a valid proof for $0$ or $1$. We properly define what a valid proof is in TODO:

\begin{dianabox}{\textsc{WeakBA}}
\algoHead{Weak Byzantine Agreement protocol for a party $p$ with input $\vin$}
\begin{algorithmic}
    \State $proof \gets \textsc{ProofCreation}(\vin)$
    \State $proofs \gets \textsc{ValidatedDispersal}(proof)$
    \If{$proofs$ only contains proofs for a single value $v \in \{0,1\}$}
        \State Decide $v$
    \Else
        \State Decide $\bot$
    \EndIf
\end{algorithmic}
\end{dianabox}

\begin{theorem}
    \textsc{WeakBA} is a weak agreement protocol.
\end{theorem}

\begin{proof}
    We note that \textsc{ProofCreation} ensures that at least $n/10$ honest parties get a valid proof, and the remaining ones output a $\bot$. Therefore, at least $n/10$ parties start the validated dispersal protocol with a valid input while the remaining ones have input $\bot$, so the input conditions for the validated dispersal algorithm are respected.
    
    Termination follows from \textsc{ProofCreation}'s and \textsc{ValidatedDispersal}'s termination. 
    
    If all honest parties have the same input $b$, then according to the definition of a proof (\cref{def:proof}), a proof for $1-b$ cannot exist, so any proof obtained can only be for $b$. Because validated dispersal's coverage and validity ensure every party gets at least a proof, it must be for $b$. Hence the protocol will decide $b$ for every honest party, so validity is satisfied.

    Using validated dispersal's coverage, there exists a proof for a value $b \in \{0,1\}$ that every honest party receives. We note this implies that parties can only decide $b$ or $\bot$, so weak agreement is satisfied.
\end{proof}

When considering complexity, we abbreviate \textsc{WeakBA} to \textsc{WBA}, \textsc{ProofCreation} to \textsc{PC} and \textsc{ValidatedDispersal} to \textsc{VD}. Additionally, $\bitc_\valdis(n, L)$ is the per-party bit complexity to disperse an $L$-bit message among $n$ parties using validated dispersal.

\begin{observation}\label{obs:weak-comp}
    Assuming the underlying protocols are byzantine-robust, \textsc{WeakBA} is byzantine-robust. Additionally, $\bitc_{\wba}(n) = \bitc_\proofcrea(n) + \bitc_\valdis(n, \kappa n)$ and $\roundc_{\wba}(n) = \roundc_\proofcrea(n) + \roundc_\valdis(n)$.
\end{observation}
\section{Proof Creation}

We now explain the proof creation protocol. The protocol starts, like many weak byzantine protocols, with each party sending its (signed) input to every other party. It differs greatly afterwards:
\begin{itemize}
    \item If a party does not get $t+1$ votes for $0$ or $t+1$ votes for $1$, it knows that not all honest parties have the same input. Thus, it can then support votes for both $0$ and $1$. We symbolize this by setting its input to $\bot$. As stated in the overview, this idea was already used by Constantinescu et Al. \cite{fromfewtomany} for adaptive algorithms. But their protocol is leader-based and uses threshold signatures, allowing it to be simpler compared to the approach used here.
    \item Another key idea is to always be able to extract information from byzantine parties: if byzantine parties send votes, honest parties will be able to use them and make signatures. Otherwise, honest parties will be able to identify many byzantine parties, accuse them and use these accusations to form a proof.
\end{itemize}

\subsection{Proofs}

Our construction relies on proof, which allows a party to know that a value satisfies validity without any additional communication. Usually, a proof is a single $(t+1)$-threshold signature. Because we do not have access to threshold signatures, our proofs require more insight. More specifically, we rely on votes, where a party $p$ signs a message $\textsc{Vote}(p, v)$ for a value $v \in \{0,1, \bot\}$, and accusations, where a party $p_i$ signs a message $\textsc{Accuse}(p_i, p_j)$ to accuse a party $p_j$. We consider the following properties:
\begin{itemize}
    \item \textbf{Voting accuracy}: An honest party can only vote for $v \in \{0,1\}$ if its input is $v$. Moreover, it can only vote for $\bot$ if it knows both $0$ and $1$ are valid (i.e not all honest parties have the same input).
    \item \textbf{Accusation accuracy}: If an honest party $p_i$ accuses $p_j$, then $p_j$ is byzantine.
\end{itemize}

Using votes and accusation, we can define what is a proof in the context of our algorithm:
\begin{definition}\label{def:proof-pract}
    A message $m$ is a proof for $b \in \{0,1\}$ if it is made of $k$ votes and $q$ accusations such that:
    \begin{itemize}
        \item All votes are for $b$ or $\bot$
        \item All parties voting and parties contained in each accusation (both the accuser and accused) are entirely disjoint
        \item $k \geq 1$ and $k + q = t+1$, i.e there is at least one vote and in total $t+1$ votes or accusations.
    \end{itemize}
\end{definition}

We note that using this definition, a proof has size $\Theta(\kappa (t+1)) = \Theta(\kappa n)$.

\begin{lemma}
    Assuming votes and accusations are accurate, a proof, as defined above in \cref{def:proof-pract}, satisfies the definition given in \cref{def:proof}.
\end{lemma}

\begin{proof}
    Consider a proof for $b \in \{0,1\}$. We want to show that $b$ satisfies validity. Using voting accuracy, it is enough to show that an honest party voted for $b$ or $\bot$. The proof is made from $k$ votes for $b$ or $\bot$ and $q$ accusations. Using accusation accuracy, each pair of accuser/accused contains at least one byzantine party. Because they are all disjoint, they are at least $q$ byzantine parties among the $2q$ parties that are part of accusations. Therefore, there are at most $t-q=k-1$ byzantine parties remaining. Using disjointness again and the fact that $k \geq 1$, this means that there is at least one honest party among the $k$ votes, which shows that $b$ satisfies validity.
\end{proof}

\subsection{Overall protocol}

Our protocol starts with two rounds of signing and sending your own vote to every party, then sharing votes received in the second round in a different way, as well as obtaining accusations if many byzantine parties were silent.

After the first round of sending votes, if an honest party does not receive $t+1$ votes for the same value, it sets its input to $\bot$, which is used for the second round of sending votes.

After the second round of voting, all votes received are preserved. Additionally, each party keeps track of other parties which did not send a vote in the previous round. It knows these are byzantine (as they did not follow the protocol) and thus can accuse them later on.

\begin{dianabox}{\textsc{ProofCreation}}
\algoHead{Proof creation protocol for a party $p$ with input $\vin$}
\begin{algorithmic}
    \State Send signed $\textsc{Vote}(p, \vin)$ to everyone
    \If{did not receive $t+1$ votes for $0$ or $t+1$ votes for $1$}
        \State $\vin \gets \bot$
    \EndIf
    \State Send signed $\textsc{Vote}(p, \vin)$ to everyone

    \State Let $Votes$ be the set of votes received in the second round
    \State Let $Accusers$ be the set of parties which sent a vote for $0$ or $1$ to $p$ in the second round
    \State Let $Accused$ be the set of parties which did not send a vote to $p$ in the second round
    
    \State $Accusations \gets$ \textsc{GetAccusations}($Accusers$, $Accused$)
    \State $Votes \gets $ \textsc{SpreadVotes}($Votes$)

    \If{\textsc{ComputeProof}($0, Votes, Accusations$) or \textsc{ComputeProof}($1, Votes, Accusations$) returns a proof $p$}
        \State Return $p$
    \Else{}
        \State Return $\bot$
    \EndIf{}
\end{algorithmic}
\end{dianabox}

In this whole section, let $nb_\bot$ be the number of honest parties which changed their input to $\bot$, and therefore sent a signed input for $\bot$ in the second round. We remark that if a party did not change its input to $\bot$, it implies it has already enough votes for a proof. Thus $n - t - nb_\bot \geq n/2 - nb_\bot$ parties already have enough votes for a proof, hence the following observation:

\begin{observation}
    If $nb_\bot \leq 4n/10$, then at least $n/10$ honest parties will be able to get a proof. 
\end{observation}

As a consequence, the following protocols (getting accusations and vote spreading) only need to focus on the case where $nb_\bot > 4n/10$. We now have two different cases, depending on the behavior of byzantine parties:
\begin{itemize}
    \item If at least one honest parties receives many votes from byzantine parties (more specifically, if one honest party receives at least $7n/10$ votes), then the vote spreading procedure will ensure that $4n/10$ parties will be able to receive enough votes to form a proof.
    \item Otherwise, it means every honest party is able to identify at least $3n/10$ byzantine parties. In this case, we remark that if an honest party manages to make $n/10$ honest parties with non-$\bot$ votes accuse disjoint byzantine parties, then these accusations along with the $\bot$ votes will be enough to form a proof. This is what the accusation procedure does.
\end{itemize}

\subsection{Computing accusations}

As stated above, computing accusations is only needed when $nb_\bot > 4n/10$ and each honest party is able to identify at least $3n/10$ byzantine parties. As a consequence, we ignore parties claiming to know less than $3n/10$ byzantine parties or when a party needs too many accusations (to be more specific, more than $3n/10$). 
The protocol consists of a leader, asking queries to other parties acting as helpers. It has $\mathcal{O}(n \log^2 n)$ bit complexity, $\mathcal{O}(\log n)$ round complexity and allows the leader to get its desired matching. We run it in parallel $n$ times, with each party acting as the leader of its own instance.

Because we have a set of at most $3n/10$ entities, each interested in at least $3n/10$ parts of a common set of size $n$ (the parties), we can represent this as a fair cake cutting problem: the cake is the set of parties and each accuser wants a piece of the cake which contains a party to accuse. We adapt the divide and conquer approach described by Even and Paz \cite{cakecutting} to the distributed and online setting. By repeatedly asking helpers (accusers) to split a range of the cake (parties) into two equal parts, we can obtain a fair cake cutting with significantly less communication.

\begin{dianabox}{\textsc{GetAccusations}($Accusers$, $Accused$)}
\algoHead{Return a set of accusations for a party $p$}
\begin{algorithmic}
    \State Send $|Accused|$ to everyone
    \State Let $A$ be the parties in $Accusers$ which accuse at least $3n/10$ parties
    \State $M \gets \emptyset$

    \InParallel{for $i = 1, \ldots,n$}
        \If{$p = p_i$ and $|A| \leq 3n/10$}
            \State $M \gets \textsc{CakeLeader}(A, 1, n)$, in parallel
        \EndIf
        \State Call $\textsc{CakeHelper}(p_i, Accused)$, in parallel
    \EndParallel{}
    \State Return $M$
\end{algorithmic}
\end{dianabox}

We now describe the behavior of a helper party. Given the set of byzantine parties $B$ it knows, it can only answer two kinds of queries:
\begin{itemize}
    \item At most $\lceil \log n\rceil$ split queries: given an interval of parties $\{p_l, p_{l+1}, \ldots, p_r \}$, and a position $i$, it returns the index of the $i$-th party in $B \cap \{p_l, p_{l+1}, \ldots, p_r \}$.
    \item At most a single accusation query: given a range of parties, it accuses one of them if it is in $B$.
\end{itemize}

\begin{dianabox}{\textsc{CakeHelper}($p_i$, $B$)}
\algoHead{Help a party $p_i$ get accusations by answering queries, for a party $p$}
\begin{algorithmic}
    \Upon{Receiving $(Accuse, l, r)$ from $p_i$}
        \If{there exists $p_j \in B \cap \{p_l, p_{l+1}, \ldots , p_r\}$}
            \State Send signed $\textsc{Accusation}(p, p_j)$ to $p_i$
        \EndIf{}
        \State Stay silent for the rest of the subprotocol
    \EndUpon

    \For{$2\lceil \log n\rceil$ rounds}
        \Upon{Receiving $(SPLIT, l, r, i)$ from $p_i$}
            \State $Q \gets \{p_l, p_{l+1}, \ldots , p_r\} \cap B$
            \State Let $p_j$ be the $i$-th party in $Q$ if it exists, otherwise $j = 0$
            \State Send $j$ to $p_i$
            \State Wait $2$ rounds
        \EndUpon
    \EndFor{}
\end{algorithmic}
\end{dianabox}

We can finally describe the protocol used by the leader. This protocol has input a set $A$ of accusers and a range parametrized by two integers $1 \leq l \leq r \leq n$. Let $\mathcal{I} = [l, l+1, ..., r]$ and $h$ the number of honest parties in $A$. We show that this protocol returns a disjoint matching of size at least $h$ between parties in $A$ and parties in $\{p_l, p_{l+1}, \ldots , p_r\}$. To do so, this algorithm divides $\mathcal{I}$ into two halves of the same size, then using helper queries, is able to partition $A$ into these two segments and recurse on them. 

\begin{dianabox}{\textsc{CakeLeader}($A$, $l$, $r$)}
\algoHead{Return a set of accusations from $A$ to $\{p_l, \ldots , p_r\}$}
\begin{algorithmic}
    \If{$|A| = 1$}
            \State Send $(Accuse, l, r)$ to the party in $A$
            \State Wait two rounds and return the set of accusations received
    \Else
        \State Send $(SPLIT, l, m, \lceil |A|/2 \rceil)$ to every party in $A$
        \State Wait two rounds for answers $(a_p)_{p\in A}$ ($a_p = 0$ if $p$ didn't answer a positive number)
        \State Let $A_l$ be the $\lceil |A|/2 \rceil$ parties in $A$ with lowest value $(a_q)$ (breaking ties arbitrarily)
        \State Let $m = \max_{p \in A_l} a_p$
        \InParallel{}
            \State $M_l \gets \textsc{CakeLeader}(A_l, l, m)$
            \State $M_r \gets \textsc{CakeLeader}(A \setminus A_l, m+1, r)$
        \EndParallel{}
        \State Return $M_l \cup M_r$
    \EndIf
\end{algorithmic}
\end{dianabox}

The main reason this protocol achieves its goal is because it is able to preserve the following invariant:
\begin{lemma}
    For a party $p_i$, let $B_i$ be the set of parties it can accuse. Then, for any call to $\textsc{CakeLeader}$ with an honest leader and input $A, l, r$:
    $$ \forall p_i \in A, \text{$p_i$ is honest} \Rightarrow |B_i \cap \{p_l, p_{l+1}, \ldots , p_r\}| \geq |A| $$
\end{lemma}

\begin{proof}
    We first consider the first instance where $\textsc{GetAccusations}$ calls $\textsc{CakeLeader}$. We remark that it is only called if the leader has at most $3n/10$ accusers who can accuse at least $3n/10$ parties. Additionally, it is called with $l = 1$ and $r= n$ which is the entire set of parties, so in this case:
    $$|B_i \cap \{p_l, p_{l+1}, \ldots , p_r\}| = |Bi| \geq 3n/10 \geq |A|$$
    So the invariant holds.

    Consider now a call to $\textsc{CakeLeader}$ with input $A, l, r$, where the invariant already holds. We want to show that it is still the case in the recursive calls. 
    Consider a party $p_i \in A$, which is honest. Using the invariant, $|B_i \cap \{p_l, p_{l+1}, \ldots , p_r\}| \geq |A|$. The leader sends a query for the $\lceil |A|/2 \rceil$-th party in the previous set. Because $|\lceil |A|/2 \rceil \leq |A|$, $p_i$ will answer with such a party. Let $k$ be the index of this party. We therefore have:
    \begin{align*}
        |B_i \cap \{p_l, p_{l+1}, \ldots , p_k\}| &= \lceil |A|/2 \rceil \\
        \text{and} \ |B_i \cap \{p_{k+1}, p_{k+2}, \ldots , p_r\}| &\geq |A| - \lceil |A|/2 \rceil
    \end{align*}
    We now have two cases:
    \begin{itemize}
        \item $p_i \in A_l$, then in this case $m = \max_{p \in A_l} a_p \geq a_i = k$, so the recursive call on $p_i$ will be made on a set of size $\lceil |A|/2 \rceil$ and range $[l, m]$, which contains $[l,k]$, for which $p_i$ can accuse at least $\lceil |A|/2 \rceil$ parties.
        \item $p_i \notin A_l$, then by minimality, $m \leq k$, so the recursive call on $p_i$ will be made on a set of size $|A| - \lceil |A|/2 \rceil$ and range $[m+1, r]$, which contains $[k+1,r]$, for which $p_i$ can accuse at least $|A| - \lceil |A|/2 \rceil$ parties.
    \end{itemize}
    In both cases, $p_i$ can accuse at least as many parties in the given range as the size of the new set $A$. As a consequence, by induction, the invariant is always satisfied.
\end{proof}

Another key point is that accusations are accurate:
\begin{lemma}
    Accuracy: All accusations returned by \textsc{CakeLeader} are accurate: an honest party will only accuse byzantine parties.
\end{lemma}
\begin{proof}
    Accusations returned by \textsc{MatchingLeader} come from \textsc{CakeHelper}, which themselves come from the set $Accused$ of parties which did not send a vote in the second round of the \textsc{ProofCreation} protocol. Because any party following the protocol will send a vote in the second round, it implies that these parties are byzantine. 

    We also remark that accusations are signed by the accuser, as such a byzantine party cannot generate an accusation made from an honest party if it does not already have it.
\end{proof}

Using this invariant and accusation accuracy, we can prove the main property of our matching protocols, which is that the accusations produced by honest parties form a disjoint matching.
\begin{lemma}\label{lemma:accusation-disjoint-matching}
    Consider an honest party, where \textsc{GetAccusations} calls \textsc{CakeLeader} with a set of accusers $Acc$. Let $Acc_h$ be the honest parties in $Acc$, then the set of accusations returned by \textsc{GetAccusations} contains a disjoint matching of size $|Acc_h|$, where each edge is made from an accuser in $Acc_h$ and an accused party being byzantine.
\end{lemma}

\begin{proof}
    We first note that \textsc{CakeLeader} partitions accusers into two sets and recursively calls itself on these two sets. No accuser gets "lost" between recursive calls and therefore every accuser will end up in exactly one case where $|A| = 1$.

    Let $q \in Acc_h$. Consider the call to \textsc{CakeLeader} with set $A$ such that $A = \{q\}$. Let $l,r$ be the range parameters of the protocol at that point. We remark that using the invariant, $|B \cap \{p_l, \ldots p_r\}| \geq |A| = 1$ . As such, $q$ will be able to find a party to accuse in $\{p_l, \ldots p_r\}$, which won't ask any other party for an accusation in this range. As a consequence, every honest party in $Acc$ will send an accusation to the leader, they will all be distinct, and using accusation accuracy, they will all target byzantine parties. Finally, looking at these accusations, because the accusers are the parties in $A_h$ and thus honest while all accused parties are all byzantine and distinct, all these accusations are fully disjoint.
\end{proof}

\begin{lemma}\label{lemma:accusations-complexity}
    \textsc{GetAccusations} has per-party bit complexity $\mathcal{O}( n^2 (\log^2 n + \kappa))$ and round complexity $\mathcal{O}(\log n)$. Moreover, it is byzantine-robust.
\end{lemma}

\begin{proof}
    \textsc{MatchingLeader} cuts into half the number of accusers at each recursion level, starting with at most $3n/10$ parties, thus its recursion depth is $\lceil \log (3n/10) \rceil$ because of the check in \textsc{GetAccusations}. In each recursion level, the leader sends a query to each party in its original accuser set, which has size at most $3n/10$. The query being a tag and a range made of two integers between $1$ and $n$, the overall bit complexity for the leader will be $\mathcal{O}(n \log^2 n)$. And the leader only spending two rounds before the parallel recursive calls means the round complexity is $2 \lceil \log n \rceil$. Looking at an helper party, for a specific leader, it will respond to at most $\lceil \log n \rceil$ queries and send at most one accusation. Because there are $n$ helper for a given leader, this amounts in total to $\mathcal{O}(\log^2 n + \kappa)$ bits with the same round complexity.

    \textsc{GetAccusations} makes at most $n$ parallel matching leader calls, one for each party. So, the total per-party bit complexity is $\mathcal{O}( n (\log^2 n + \kappa))$ and round complexity $\mathcal{O}(\log n)$.

    Finally, we remark that we did not use the resiliency $t < n/2$ at any point for proofs related to this protocol. As a matter of fact, it supports an arbitrary number of byzantine parties. Thus it is byzantine-robust.
\end{proof}

\subsection{Spreading votes}

The last case to handle happens when at least one honest party is not able to identify many byzantine parties. More specifically, if at least one honest party $p$ is unable to identify at least $3n/10$ byzantine parties. For this to happen, most byzantine parties must appear as following the protocol for $p$ and therefore send a vote (for $0$, $1$ or $\bot$) to it. As a consequence, in this case, $p$ will receive at least $7n/10$ votes. As a reminder, we can also assume that $nb_\bot$ (the number of $\bot$ votes by honest parties) is at least $4n/10$. We remark that this is enough for $p$ to get a proof: in addition to $4n/10$ $\bot$ votes, either $0$ or $1$ will get at least $(7n/10 - 4n/10)/2$ votes supporting it, which in total is strictly more than $0.55n > n/2 > t$ votes. 

However, getting a single party to get a proof is not enough, we want at least $n/10$ honest parties to get one. To do so, we remark that $p$ has a "surplus" of $0.05$ votes. So, by forwarding votes to other parties using a disperser pattern, which only requires parties to forward each vote to a constant number of parties, we can make it so that any subset of $4n/10$ parties has enough votes together to get a proof. A key insight is that we forward each vote to a different set of parties, so even if a single vote may only be forwarded to byzantine parties, only an arbitrary small fraction of the votes will. Finally, using a fault-proof expander graph, we can find a subset of $4n/10$ honest parties which are all at logarithmic distance of each other and such that the graph degree is constant.

\begin{dianabox}{\textsc{SpreadVotes}($Votes$)}
\algoHead{Spread votes among all parties for a party $p$}
\begin{algorithmic}
    \State Let $G_1$ be a $(n/20, 7n/10)$-disperser
    \Comment Using \cref{theo:big-disperser} with $\alpha = 1/20$ and $m = 115$
    \For{each vote $v$ from party $q$ in $Votes$, in parallel}
        \State Send $v$ to $N_{G_1}(q)$
    \EndFor{}

    \State Let $G_2$ be a fault-tolerant expander
    \Comment Using \cref{theo:fault-tol-exp}
    \State Send $Votes$ to $N_{G_2}(p)$
    \For{$\mathcal{O}(\log n)$ rounds}
        \Upon{Receiving a vote $v$ not in $Votes$}
            \State Forward $v$ to parties in $N_{G_2}(p)$
            \State $Votes \gets Votes \cup \{v\}$
        \EndUpon
    \EndFor{}

    \State Return $Votes$
\end{algorithmic}
\end{dianabox}

\begin{lemma}\label{lemma:spread-complexity}
    \textsc{SpreadVotes} has per-party bit complexity $\mathcal{O}(n \kappa)$ and round complexity $\mathcal{O}(\log n)$. Additionally, it is byzantine-robust.
\end{lemma}

\begin{proof}
    The graphs ($G_1$ and $G_2$) used in the protocol using \cref{theo:big-disperser,theo:fault-tol-exp} have constant degree. We note that there can only exist $3n$ votes in total: $n$ parties and $3$ votes ($0$, $1$ or $\bot$) per party, and each vote is $\mathcal{O}(\kappa)$ bits. Because each party sends each unique vote to at most $\mathcal{O}(1)$ parties using $G_1$ and $G_2$, the total per-party bit complexity is $\mathcal{O}(n \kappa)$.
    Regarding round complexity, the first step takes $1$ round while the second step takes $\mathcal{O}(\log n)$ rounds. So the overall round complexity is $\mathcal{O}(\log n)$ round. Finally, we remark that the resiliency $t < n/2$ was not used when computing the complexity, as such this protocol is byzantine-robust.
\end{proof}

\begin{lemma}\label{lemma:spread-subset-large}
    Consider a party $p$ and a subset $S$ of its votes, where each vote is from a distinct party. For any party $q$, let $E_q \subseteq S$ be the votes from $S$ it received after the first round. Then for any subset $H \subseteq \mathcal{P}$ of parties such that $|H|\geq 4n/10$, $|\cup_{q \in H} E_Q| \geq |S| - n/20$. 
\end{lemma}

\begin{proof}
    Consider a party $p$ and sets $S$ and $H$ as defined in the lemma. Let $R$ be the set of votes from $S$ that were received by no parties in $H$, i.e $R = S \setminus \cup_{q \in H} E_Q$. We want to prove that $|R| \leq n/20$. Because each vote in $S$ comes from a different party, we can identify votes with the party that emitted it. Because votes were transmitted using the graph $G_1$, this implies that there is no edge between $R$ and $H$, so $N_{G_1}(R) \subseteq \mathcal{P} \setminus H$, so $|N_{G_1}(R)| \leq n - |H| \leq n - 4n/10 \leq 6n/10$. $G_1$ is a disperser, any set of size $n/20$ has neighborhood of size at least $7n/10$. By contrapositive, $|R| < n/20$.
\end{proof}

\begin{lemma}\label{lemma:spread-form-proof}
    If $nb_\bot \geq 4n/10$ and an honest party has votes from at least $7n/10$ parties, then at least $4n/10$ parties will receive enough votes to form a proof.
\end{lemma}

\begin{proof}
    Assume $nb_\bot \geq 4n/10$ and an honest party $p$ has votes from more than $7n/10$ honest party. Let $R$ be the set of votes $p$ has, keeping only one per party. We have $|R| \geq 7n/10$. Because we only consider one vote per party, we can identify a vote with the party that emitted it.

    Let $Bz$ be the set of byzantine parties, we have $|Bz| \leq t \leq n/2$. Using the property on $G_2$ (\cref{theo:fault-tol-exp}), there exists a set of parties $S$ such that $S$ and $Bz$ are disjoint, i.e $S$ is honest, $|S| \geq 4n/10$, and $G_2[S]$ is connected and has diameter $\mathcal{O}(\log n)$. We can now use \cref{lemma:spread-subset-large} on $S$ and $R$: there exists $R' \subseteq R$ such that $|R'| \geq |R| - n/20 \geq 7n/10 - n/20 = 6.5n/10$ and parties in $S$ together receive all votes of $R'$ from $p$ in the first round of \textsc{SpreadVotes}. Therefore, after all $\mathcal{O}(\log n)$ rounds, all parties of $S$ will have all votes of $R'$. 

    We now show that votes in $R'$ are enough to form a proof. Because $nb_\bot \geq 4n/10$, there are at least $4n/10$ votes for $\bot$ in $R$. Since $R' \subseteq R$ and $|R'| \geq |R| - n/20$, $R'$ has at least $4n/10 - n/20 = 3.5n/10$ votes for $\bot$. Let $x$ be the number of votes for $\bot$ in $R'$ and let $y$ be the number of votes for $0$ or $1$. We have $x + y = |R'| \geq 6.5n/10$. One of $0$ and $1$ has at least $y/2$ votes in $R'$. Hence, the votes for $\bot$ together with the votes for this value are at least:
    \begin{align*}
        x + \frac{y}{2} &= \frac{x + (x+y)}{2} \\
        &= \frac{x + |R'|}{2}\\
        &\geq \frac{3.5n + 6.5n}{2} \\
        &= n/2
    \end{align*}
    
    Since $t<n/2$, there are at least $t+1$ such votes, all from distinct parties, so a proof can be formed. Thus, all parties in $S$ will be able to get form a proof and $|S| \geq 4n/10$, which concludes the proof.
\end{proof}

\subsection{Computing a Proof}

We finally describe the last step of the protocol, which is to compute a proof from votes and accusations. This is done by computing a maximum matching on accusations where neither accuser nor accused can be used as votes.

\begin{dianabox}{\textsc{ComputeProof}($b, Votes, Accusations$)}
\algoHead{Compute a proof for $b$ if one can be found}
\begin{algorithmic}
    \State Let $Q$ be the votes in $Votes$ for $v$ or $\bot$
    \State Let $E$ be the $Accusations$ disjoint from $Q$, seen as edges
    \State Compute a maximum matching $M$ of $E$
    \If{$|Q| \geq 1$ and $|Q| + |M| \geq t+1$}
        \State Return a set of $t+1$ votes/accusations from $Q$ and $M$ with at least one vote
    \Else 
        \State Return $\bot$
    \EndIf
\end{algorithmic}
\end{dianabox}

The maximum matching can be computed in polynomial time using the Blossom Algorithm. As a side note, looking at how we got accusations, we remark that the undirected graph the matching is computed on comes from a directed graph where each vertex has in-degree and out-degree at most $1$. Hence, it is a collection of disjoint paths and cycles, which allows a maximum matching to be computed in linear time.

\begin{lemma}\label{lemma:compute-proof-good}
    If there exists a proof for $b$ using $Votes$ and $Accusations$, then a call to \textsc{ComputeProof} will return a proof for $b$.
\end{lemma}

\begin{proof}
    Consider a proof for $b$, let $Q$ be its set of votes and $T$ its set of accusations. We note that, without loss of generality, we can assume that no party present in $T$ (as an accuser or accused) has a vote for $b$ or $\bot$ in $Votes$. The reason for it is that if this were the case, we could remove the accusation from $T$ and add the vote to $Q$, which would result in a valid proof. 

    Let $V_b$ be the set of votes in $Votes$ for $b$ or $\bot$ and let $E_b$ be the set of accusations in $Accusations$ disjoint from $V_b$. We have $Q \subseteq V_b$ and, because no accuser or accused in $T$ has a vote for $b$ or $\bot$, $T \subseteq E_b$. As a consequence, $|V_b| \geq |Q| \geq 1$. Moreover, $T$ is a matching of $E_b$, so there exists a matching of size $|T|$. By maximality of $M$, we will have $|M| \geq |T|$. Therefore, $|V_b| + |M| \geq |Q| + |T| = t+1$, which shows that \textsc{ComputeProof}($b, Votes, Accusations$) will return a proof for $b$.
\end{proof}

\subsection{Putting it together}

\begin{theorem}
    After running \textsc{ProofCreation}, at least $n/10$ honest parties will have a proof for $0$ or $1$. 
\end{theorem}

\begin{proof}
    We remind the reader that $nb_\bot$ is the number of honest parties which changed their input to $\bot$. After the first round of sending votes. We now have $3$ different cases to consider:
    \begin{itemize}
        \item $nb_\bot \leq 4n/10$: If a party does not change its input to $\bot$, it means it either received $t+1$ votes for $0$ or $t+1$ votes for $1$. So, it has already enough votes to form a proof. Because $t < n/2$, at least $n - t - nb_\bot \geq n/2 - 4n/10 = n/10$ honest parties will have enough votes to form a proof.
        \item $nb_\bot > 4n/10$ and at least one honest party received votes from at least $7n/10$ parties: Using \cref{lemma:spread-form-proof}, at least $4n/10$ parties will receive enough votes to form a proof.
        \item $nb_\bot > 4n/10$ and every honest party received less than $7n/10$ votes: In this case, every honest party is able to identify at least $3n/10$ byzantine parties. Moreover, given an honest party $p$, its set of accusers $Acc$ is made of parties which sent a vote to $p$ different than $\bot$. So $|Acc| \leq n - 3n/10 - nb_\bot \leq 3n/10$. As a consequence, \textsc{GetAccusations} will call \textsc{MatchingLeader} for every honest party. Consider an honest party $p$, either $p$ is counted towards $nb_\bot$ and sent a $\bot$-vote to everyone, or it sent a vote for $0$ or $1$ to everyone, which implies it is in the set of accusers of every other honest party. Using \cref{lemma:accusation-disjoint-matching}, the set of accusations returned by \textsc{GetAccusations} contains a disjoint matching between all such honest parties as accusers and byzantine parties as accused. Thus, these accusations are disjoint from $\bot$-votes of the $nb_\bot$ honest parties. Because each honest party appears exactly once with a $\bot$-vote or as an accuser, we have a set of votes and accusations of size $n - t \geq t+1$ with at least one vote ($nb_\bot > 0$), which is enough to form a proof. So all honest parties will be able to form a proof.
    \end{itemize}
    Finally, using \cref{lemma:compute-proof-good}, the $n/10$ honest parties that have enough votes and accusations will return a proof for $0$ or $1$.
\end{proof}

\begin{theorem}\label{theo:proof-crea-comp}
    \textsc{ProofCreation} has bit complexity $\bitc_{\proofcrea}(n) = \mathcal{O}(n^2 (\log^2 n + \kappa))$ and round complexity $\roundc_{\proofcrea}(n) = \mathcal{O}(\log n)$. Additionally, it is byzantine-robust.
\end{theorem}

\begin{proof}
    The first two rounds of \textsc{ProofCreation} have bit complexity $\mathcal{O}(n^2 \kappa)$, as each party sends a vote to every other party. Combining it with \cref{lemma:spread-complexity,lemma:accusations-complexity}, we get the expected complexities.
\end{proof}
\section{Validated dispersion}

We now consider the second part of our protocol, which is validated dispersion. After the first part, $n/10$ honest parties are guaranteed to hold a proof of size $\mathcal{O}(n \kappa)$. The goal of validated dispersion is to ensure that there is one proof which ends up known by all honest parties. We note that a very simple implementation of validated dispersion is to have each party send its proof to all other parties if it has one. Although it is quite simple, the total bit complexity would be $\mathcal{O}(n^3 \kappa)$.

Our protocol instead relies on partitioning all $n$ parties into $m$ groups of size $\lfloor n/m \rfloor$ or $\lceil n/m \rceil$, where $m := m(n)$ is a parameter to be determined later. We first remark that, by sending proofs following a disperser graph, we can ensure that all but at most $n/(10m)$ honest parties will have a proof. This has bit complexity $\mathcal{O}(n^2 m \kappa)$. Afterwards, we can agree on a proof in each group and broadcast it efficiently using erasure codes. Because one of the groups is guaranteed to have a byzantine minority, this ensures that the proof from this group will reach every party. In the end, the protocol returns the set of proofs received from each group.

The group broadcast protocol is described in \cref{subsec:group-broadcast}. An assumption we make is that $m = \mathcal{O}(n^\alpha)$ for some $0 < \alpha < 1$ as a way to simplify the complexity analysis. We show that this assumption is satisfied in \cref{sec:together}.

\begin{dianabox}{\textsc{ValidatedDispersion}($proof$)}
\algoHead{Disperse a proof for a party $p$}
\begin{algorithmic}
    \State $proofs \gets \emptyset$
    \State Let $G$ be a $(n/10, n - n/(10m))$-disperser
    \Comment{Using \cref{theo:big-disperser} with $\alpha = 1/10$ and degree $4 * 10 * 10 * m$}
    \State Send $proof$ to $N_G(p)$
    \If{$proof = \bot$ and received a valid proof $q$}
        \State $proof \gets q$
    \EndIf{}

    \State Partition $\mathcal{P}$ into $m$ groups of size $\lfloor n/m \rfloor$ or $\lceil n/m \rceil$.
    \For{each group $G$ in parallel}
        \If{$p \in G$}
            \State $q \gets \textsc{PartialValidatedAgreement}(G)$
        \EndIf
        \State $q \gets \textsc{GroupBroadcast}(G, q)$
        \If{$q$ is valid}
            \State $proofs \gets proofs \cup \{q\}$
        \EndIf
    \EndFor

    \State Return $proofs$
\end{algorithmic}
\end{dianabox}

\begin{theorem}
    The protocol satisfies validated dispersion.
\end{theorem}

\begin{proof}
    Termination comes from \textsc{PartialValidatedAgreement} and \textsc{GroupBroadcast}'s termination. Because a value only gets added to $proofs$ if it is valid, validity follows.

    We assume that at least $n/10$ honest parties have a non-$\bot$ valid input. Because we send values using a $(n/10, n - n/(10m))$-disperser, at most $n/(10m) \leq \lfloor n/m \rfloor/10$ honest parties will not receive a valid input. Parties are partitioned into groups of size at least $\lfloor n/m \rfloor/10$ . Because the initial set of parties has a byzantine minority, it implies that at least one of the groups has a byzantine minority too. Consider this group $G$, of size $k$. It has a byzantine minority, and among the honest parties, at most $k/10$ do not have a valid input, so at least $k - k/10 - k/2 \geq k/3$ honest parties have a valid input.

    Therefore, using partial validated agreement, all honest parties in $G$ will agree on a valid proof $q$. Again, because $G$ has an honest majority and all honest parties call it with the same input $q$, all honest parties in $\mathcal{P}$ will receive $q$. So coverage is also satisfied.
\end{proof}

\begin{theorem}\label{theo:vd-comp}
    Assuming the partial validated agreement protocol is byzantine-robust, \textsc{ValidatedDispersion} has per-party bit complexity $\bitc_{\valdis}(n, L) = \mathcal{O}(m L ) + \bitc_{\textsc{PVA}}(\lceil n/m \rceil, L)$, round complexity $\roundc_{\valdis}(n) = \mathcal{O}(1) + \roundc_{\textsc{PVA}}(\lceil n/m \rceil, L)$ and is byzantine-robust.
\end{theorem}

\begin{proof}
    The disperser $G$ is obtained from \cref{theo:big-disperser} using degree $4 * 10 * 10 * m = \mathcal{O}(m)$, so each party sends its proof of size $L$ to $\mathcal{O}(m)$ parties, which gives a per-party bit complexity of $\mathcal{O}(m L)$. Afterwards, each party calls \textsc{PartialValidatedAgreement} on a group of size at most $\lceil n/m \rceil$, which gives a per-party bit complexity of $\bitc_{\textsc{PVA}}(\lceil n/m \rceil, L)$. Finally, the group broadcast has cost $\mathcal{O}(L n / \lceil n/m \rceil + \kappa n \log n) = \mathcal{O}(m L)$ as we assume $m = \mathcal{O}(n^\alpha)$ for $\alpha > 0$. Regarding the round complexity, every part of this protocol except the partial validated agreement takes $\mathcal{O}(1)$ rounds.
\end{proof}
\section{Parallel agreement}\label{sec:parallel-agreement}

Using validated dispersion, we go from solving a problem on $n$ parties to solving it on $m << n$ parties. A consequence, is that we can recursively run our own byzantine agreement protocol on this smaller set of parties. However, this agreement protocol is still expensive. We remark that we mostly use parallel reliable broadcast in our partial validated agreement protocol, and as such, optimizing this use case would improve the overall complexity of our protocol. To do so, we rely on batching, where given a batch size $b$, we only need to run recursively our byzantine agreement $\kappa$ times for $b$ parallel invocations. 

\subsection{Committee Broadcast}

We first describe our committee broadcast protocol. This is a variant of the Dolev-Strong protocol, where only a smaller committee of size $b$ is responsible for forwarding signatures, instead of all parties. The advantage of such approach is that it has lower complexity than the original Dolev-Strong protocol. The disadvantage is that agreement is only guaranteed if at least one honest party in the committee is honest.

Our broadcast protocol is used to send a small value. More specifically, we assume the input $\vin$ satisfies $|\vin| = \mathcal{O}(\kappa)$.

\begin{dianabox}{\textsc{CommitteeBroadcast}($\vin, p_{in}, C$)}
\algoHead{Code for a party $p_{in} \in C$ to broadcast a value $\vin$, executed by a party $p$}
\begin{algorithmic}
    \If{$p = p_{in}$}
        \State Sign $\vin$ and send it to all parties
    \EndIf
    \For{$i = 2$ to $|C|+1$}
        \Comment{An iteration lasts $2$ rounds}
        \Upon{receiving a message signed by $p_{in}$ and $i-1$ parties from $C$}
            \State Forward it to all parties in $C$
            \Comment{Only applied for two distinct messages received}
        \EndUpon
        \If{$p \in C$}
            \Upon{receiving a message signed by $p_{in}$ and $i-1$ parties from $C$}
                \State Add your signature and forward it to all parties
                \Comment{Only applied for two distinct messages received}
            \EndUpon
        \EndIf
    \EndFor
    \If{received a single message $m$}
        \State Return $m$
    \Else
        \State Return $\bot$
    \EndIf
\end{algorithmic}
\end{dianabox}

\begin{lemma}\label{lemma:committee-ba-prop}
    If at least one party in $C$ is honest, then the protocol satisfies byzantine broadcast.
\end{lemma}

\begin{proof}
    Assume there exists a party $q \in C$ that is honest.

    We first show that agreement is satisfied. We remark that if an honest party in $C$ forwards $2$ distinct signed messages from the sender, every party will receive these messages and decide $\bot$, in which case agreement is satisfied. Otherwise, consider the case where honest parties in $C$ only forward at most one signed message. Consider any honest party $p \in \mathcal{P}$ that accepts a signed message $m$, we want to show that every other honest party will receive $m$. We have two cases:
    \begin{itemize}
        \item Either the message was signed by $q$. This implies that $q$ signed it in a previous iteration, and forwarded it to all parties, so every honest party will receive it.
        \item Or the message was not signed by $q$. We remark in this case that we are not in iteration $|C|+1$, as otherwise the message would have to be signed by all parties in $C$, including $q$. In this case, $p$ will forward $m$ to all parties in $C$ including $q$. Because there is at least one iteration remaining, $q$ will add its signature and forward it to all parties, so every honest party will receive and accept it.
    \end{itemize}
    As a consequence, the set of messages accepted by all honest parties is the same, so the decided value will be the same: agreement is satisfied.
    Given that the protocol runs for $\mathcal{O}(|C|)$ rounds and then returns a value, termination is satisfied. Finally, assume the sender $p_{in}$ is honest. Then, it will sign and send the same input $\vin$ to all parties, which will receive and accept it. Moreover, $p_{in}$ being honest will not sign any other message, so no other message will be accepted. As a consequence, all honest parties will decide $\vin$, so validity is satisfied.
\end{proof}

\begin{lemma}\label{lemma:committee-comp}
    The protocol has round complexity $\mathcal{O}(|C|)$. Regardless of messages received, parties in $C$ send $\mathcal{O}( \kappa n |C|)$ bits while other parties send $\mathcal{O}( \kappa|C|^2)$bits.
\end{lemma}

\begin{proof}
    We note that the protocol runs for exactly $2|C| + 1$ rounds. All messages contain an input, which is $\mathcal{O}(\kappa)$ bits and at most $|C|$ signatures, which are each $\kappa$ bits. Parties in $C$ send at most $2$ such messages to each party, while other parties send at most $2$ messages to parties in $C$, hence the result.
\end{proof}

\subsection{Batch Broadcast}

Our previous committee broadcast is reliable as long as at least one party in the committee is honest. We use it to implement batch broadcast, where a group of $C$ parties want to broadcast their $L$-bit input at the same time. We remind the reader that we assume $L = \mathcal{O}(\kappa)$. We remark that if we use the previous committee broadcast with $C$ being the committee, then either at least one party in $C$ is honest, in which case we are done, or all parties in $C$ are byzantine. In the latter case, we only care about agreement, so we can run a single byzantine agreement on the concatenated output of all broadcast result. This approach allows us to reduce the number of byzantine agreement calls from $|C|$ to $1$, which is a significant improvement. The \textsc{MVBA} protocol is described in \cref{subsec:mvba}.

\begin{dianabox}{\textsc{BatchBroadcast}($\vin, C$)}
\algoHead{Code for all parties in $C$ to broadcast an input $\vin$, executed by a party $p$}
\begin{algorithmic}
    \InParallel{For each party $q \in C$}
        \State Call $\textsc{CommitteeBroadcast}(\vin, q, C)$
    \EndParallel
    \State Let $m$ be the concatenation of all outputs from the previous calls
    \State Return \textsc{MVBA}($m$)
\end{algorithmic}
\end{dianabox}

\begin{lemma}\label{lemma:batch-broad-prop}
    The protocol satisfies byzantine broadcast for each party in $C$.
\end{lemma}

\begin{proof}
    Termination follows from \textsc{CommitteeBroadcast} and \textsc{MVBA}'s termination. 
    We consider validity. We remark that validity is only relevant if at least one party in $C$ is honest. But in this case, using \cref{lemma:committee-ba-prop}, the output of each committee broadcast satisfies validity and agreement. As a consequence, each honest party will receive the same output for each broadcast, which is the parties' input if it is honest. Thus, the concatenated output will be the same for all honest parties, and \textsc{MVBA}'s validity will ensure that the final output is the same as the common input. So every honest party will output the parties' input if it is honest, and validity is satisfied. 

    Finally, agreement is a consequence of \textsc{MVBA}'s agreement.
\end{proof}

\begin{lemma}\label{lemma:batch-broad-comp}
    The protocol has round complexity $\mathcal{O}(|C|) + \roundc_{\textsc{MVBA}}(n)$. Regardless of messages received, parties in $C$ send $\mathcal{O}(\kappa n |C|^2) +  \bitc_{\textsc{MVBA}}(n, |C| L)$ bits while other parties send $\mathcal{O}(\kappa |C|^3 ) + \bitc_{\textsc{MVBA}}(n, |C| L)$ bits.
\end{lemma}

\begin{proof}
    This is a direct application of \cref{lemma:committee-comp} and the fact that the concatenation of all outputs is of size $|C| L$.
\end{proof}

\subsection{Parallel broadcast}

We now describe our final protocol for parallel byzantine agreement. The idea is to partition the set of parties into groups of size $b := b(n)$, and run \textsc{BatchBroadcast} on each group in parallel. This allows us to reduce the number of byzantine agreement calls from $n$ to $\lceil n/b \rceil$. We note that there is a trade-off on $b$: the larger $b$ is, the less amount of recursive byzantine agreement calls we need to make, but at the same time the more expensive the \textsc{CommitteeBroadcast} becomes. The exact choice is left for later, as it is chosen to minimize the bit complexity of the overall byzantine agreement protocol.

\begin{dianabox}{\textsc{ParallelBroadcast}($\vin, C$)}
\algoHead{Code for all parties in $C$ to broadcast an input $\vin$, executed by a party $p$}
\begin{algorithmic}
    \State Partition $C$ into groups of size $b$ (except possibly the last group)
    \For{each group $C_i$ in parallel}
        \State Call \textsc{BatchBroadcast}($\vin, C_i$)
    \EndFor
    \State Return the output of all previous calls
\end{algorithmic}
\end{dianabox}

\begin{lemma}
    The protocol satisfies byzantine broadcast for each party in $C$.
\end{lemma}

\begin{proof}
    This is a direct consequence of \cref{lemma:batch-broad-prop}.
\end{proof}

\begin{lemma}\label{lemma:parallel-broad-comp}
    The protocol has round complexity $\mathcal{O}(|C|) + \roundc_{\mvba}(n)$. Regardless of messages received, parties send $(|C|/b + 1)\left(\mathcal{O}(\kappa b^3) +  \bitc_{\textsc{MVBA}}(n, b L)\right)$ bits. Parties in $C$ send an additional $\mathcal{O}(\kappa n b^2)$ bits.
\end{lemma}

\begin{proof}
    The round complexity follows from the fact that we only run parallel executions of \textsc{BatchBroadcast} and \cref{lemma:batch-broad-comp}.
    Regarding the bit complexity, we make exactly $\lceil |C|/b \rceil \leq |C|/b + 1$ calls to \textsc{BatchBroadcast}, and each of these calls is run with a group size $|C| \leq b$. Using \cref{lemma:batch-broad-comp}, this results in the expected bit complexity. Finally, if a party is in $C$, exactly one of the batch broadcasts will be on the party's group, which using \cref{lemma:batch-broad-comp} results in the additional $\mathcal{O}(\kappa n^2 b)$ bits sent.
\end{proof}
\section{Partial Validated Agreement}

We now describe a protocol to reduce partial validated agreement to validated agreement. In partial validated agreement, only $n/3$ honest parties are guaranteed to have a valid input. The main insight is to have each party broadcast reliably whether it has a valid input or not. Then, we run the validated agreement protocol only on parties that claim to have a valid input. Afterwards, we use a group broadcast protocol to ensure that all parties receive the same decision value. If it is not valid, this implies that the validated agreement protocol was run on a byzantine majority, so the remaining parties form a byzantine minority. Thus, we can first use a disperser graph to ensure that a third of the remaining honest parties have a valid input, then iterate back onto them.

One point to note is that we want our protocol to be byzantine-friendly, so we must not iterate too many times if this should not be possible. In particular, if less than a third of the parties claim to have a valid input, this cannot happen with a byzantine minority, so we can just stay silent for the rest of the protocol.

The \textsc{ReliableGroupBroadcast} protocol is described in \cref{subsec:rel-group-broadcast}.

\begin{dianabox}{\textsc{PartialValidatedAgreement}($\vin$)}
\algoHead{Agree on a valid input for a party $p$}
\begin{algorithmic}
    \State $Q \gets \mathcal{P}$

    \For{$\lceil \log_{3/2} n \rceil$ iterations}
        \State Let $G$ be a $(n/3, 5|Q|/6)$ bipartite disperser
        \Comment{ Using \cref{theo:small-bipartite}}
        \State Send $\vin$ to $N_G(p)$
        \If{$\vin = \bot$ and received a valid input $v$ from a neighbor}
            \State $\vin \gets v$
        \EndIf

        \State $(b_q)_{q \in Q} \gets \textsc{ParallelBroadcast}(Q, \vin \ne \bot)$
        \State $Q_1 \gets \{q \in Q : b_q = 1\}$
        \If{$|Q_1| < |Q|/3$}
            \State Stay silent for the rest of the protocol
            \Comment{Can only happen with a byzantine majority}
        \EndIf
        \If{$p \in Q_1$}
            \State $v \gets \textsc{ValidatedAgreement}_{Q_1}(\vin)$
        \EndIf
        \State $v \gets \textsc{ReliableGroupBroadcast}_Q(Q_1, v)$
        \If{$v$ is valid}
            \State Return $v$
        \EndIf
        \State $Q \gets Q \setminus Q_1$
    \EndFor
\end{algorithmic}
\end{dianabox}

\begin{theorem}
    \textsc{PartialValidatedAgreement} satisfies partial validated agreement.
\end{theorem}

\begin{proof}
    We first show by induction that, at the $i$-th iteration of the protocol, the set $Q$ has a byzantine minority and $|Q| \leq n \cdot (2/3)^{i-1}$. At first, $Q = \mathcal{P}$ which has a byzantine minority and size $n$, so initialization holds. 

    We now consider iteration $i$ and assume $Q$ has a byzantine minority and $|Q| \leq n \cdot (2/3)^{i-1}$. $G$ is a disperser used to send parties'input. Because it is a $(n/3, 5|Q|/6)$-disperser and at least $n/3$ honest parties send their valid input through it, at least $5|Q|/6$ parties receive a valid input. So at least $5|Q|/6 - |Q|/2 = |Q|/3$ parties will have a valid input. Because $Q$ has a byzantine minority, the  \textsc{ParallelBroadcast} call will succeed, and thus $|Q_1| \geq |Q|/3$. If $Q_1$ has a byzantine minority, then all the conditions for validated agreement are satisfied, so all honest parties in $Q_1$ will agree on a valid input and \textsc{ReliableGroupBroadcast} will ensure that all honest parties receive this input and return. 

    If this does not happen, then it implies that $Q_1$ has a byzantine majority, so $Q \setminus Q_1$ has a byzantine minority. Moreover, because $|Q_1| \geq |Q|/3$, we have $|Q \setminus Q_1| \leq 2|Q|/3 \leq n \cdot (2/3)^i$ using the induction hypothesis. Thus, the induction step holds.

    Finally, after $\lceil \log_{3/2} n \rceil$ iterations, if no output has been returned yet, we have $|Q| \leq n \cdot (2/3)^{\lceil \log_{3/2} n \rceil} \leq 1$. Because there is a byzantine majority, it implies that there is exactly one honest party left in $Q$. It will thus receive a valid input, share it with other parties and they will return it this iteration. Thus, all parties will return the same valid input: the protocol satisfies partial validated agreement.
\end{proof}

\begin{lemma}\label{lemma:pva-comp}
    Assuming the underlying protocols are byzantine-robust, then the protocol is byzantine-robust and:
    \begin{align*}
        \bitc_\pval(n,L) = \mathcal{O}(L \log n + \kappa n b^2 + \kappa b^3 \log n) + n/b \cdot  \bitc_\mvba(n, b) + \mathcal{O}(\log n) \cdot \bitc_{\mvba}(n, L)  \\ 
        + \sum_{i = 0}^{\lceil \log_{3/2} n \rceil}  \bitc_\fullval(n \cdot (2/3)^i, L)
        \end{align*}
        Additionally, $$\roundc_\pval(n) = \mathcal{O}(\log n) (b + \roundc_\mvba(n)) + \sum_{i = 0}^{\lceil \log_{3/2} n \rceil}  \roundc_\fullval(n \cdot (2/3)^i)$$.
\end{lemma}

\begin{proof}
    We consider the round complexity. We note that in the $i$-th iteration, an honest party will only consider that at most $(2/3)^{i-1} n$ parties are participating. However, because we want our protocol to be byzantine robust, this does not mean that at most $(2/3)^{i-1} n$ honest parties are participating: if there is a byzantine majority, then honest parties may disagree on who is participating and all of them may believe they are participating. We consider the complexity of the $i$-th iteration, while remarking that a party $p$ will consider itself to be part of at most one set $Q_1$ in the whole protocol:
    \begin{itemize}
        \item Sending using the disperser graph: using \cref{theo:small-bipartite}, $G$ has constant degree, so the per-party bit complexity is $\mathcal{O}(L)$. This takes $1$ round.
        \item Parallel broadcast: on at most $n \cdot (2/3)^{i-1}$ parties. Parties broadcast a single bit. Using \cref{lemma:parallel-broad-comp}, this takes $b + \roundc_{\mvba}(n \cdot (2/3)^{i-1})$ rounds and $\mathcal{O}(\kappa n b^2) + ((n \cdot (2/3)^{i-1})/b + 1)\cdot (\mathcal{O}(\kappa b^3) + \bitc_\mvba(n, b))$ bits per party.
        \item Validated agreement: has per-party bit complexity $\bitc_\fullval(n \cdot (2/3)^{i-1}, L)$ and round complexity $\roundc_\fullval(n \cdot (2/3)^{i-1})$.
        \item Reliable group broadcast: Using \cref{lemma:reli-group-broad-comp}, this takes $\mathcal{O}(\kappa n \log n) + \bitc_{\mvba}(n \cdot (2/3)^{i-1}, L)$ bits and $\mathcal{O}(1) + \roundc_{\mvba}(n \cdot (2/3)^{i-1})$ rounds.
    \end{itemize} 
    Summing these over all iterations, and using the fact that $\sum_{i = 0}^{\lceil \log_{3/2} n \rceil} (n \cdot (2/3)^{i-1}) = \mathcal{O}(n)$, we obtain the expected results.
\end{proof}
\section{Validated Agreement}

We now present our protocol to solve validated byzantine agreement. To do so, we first present initial ideas to perform Validated Agreement and what their limitations are.

\paragraph{Dolev-Strong:} An initial idea would be to have each party send their input using reliable broadcast and more specifically the Dolev-Strong broadcast protocol. Parties could just then pick up the first valid value they receive. This would have $\mathcal{O}(n^3(n \kappa + L))$ bit complexity. If used within our framework, this would lead to a protocol solving byzantine agreement with bit complexity $\mathcal{\tilde{O}}(n^{8/3})$. Although this is subcubic, the cubic cost of Dolev-Strong is the bottleneck in this case.

\paragraph{Recursive Application:} We remark that we can assume a byzantine minority and that byzantine broadcast can be implemented with a single call to byzantine agreement. This leads to a new idea: instead of using Dolev-Strong to implement Byzantine Broadcast, we can use our own byzantine agreement protocol recursively. Doing so, along with common erasure coding techniques, would lead to validated agreement having a bit complexity of $\mathcal{\tilde{O}}(n^2 L + n^{3.5})$. This would cause the overall bit complexity of our byzantine agreement protocol to be $\mathcal{\tilde{O}}(n^{2.5})$.  

\paragraph{Byzantine-Robust protocol:} The bottleneck of our protocol is now the $\mathcal{O}(n^2 L)$ in the bit complexity. If we ignore round complexity for a moment, a simple idea would be to perform the broadcast calls consecutively and stop as soon as one succeeds. Indeed, only the broadcaster has a $\mathcal{O}(nL)$ cost and if the broadcaster is honest, the broadcast will succeed. This idea works and decreases the $L$-dependency to $\mathcal{O}(n L)$. However, this approach does not work with our requirements: it is not byzantine-robust. If there is a byzantine majority, they can prevent our broadcast protocol from succeeding, even if the broadcaster is honest and cause the overall bit complexity to go back to quadratic in $L$. 

Our solution for these problems is to use a dynamic programming approach: we track for a given group a valid input as well as a set of parties that know this input and we show how to update this information when merging two groups. What remains is to decide how to merge groups to minimize both the round and bit complexity.

\subsection{Group Merging}\label{subsec:group-merging}

We now provide the proper invariant we maintain on each group. Given a group $G \subseteq  \mathcal{P}$, every party keeps track of the tuple:
$$ (h, H) $$
Such that $h$ is the accumulator of a valid input, all honest parties in $H$ know the value associated with $h$ and $G \setminus H$ has a byzantine majority (more specifically, at least $(|G| - |H|)/2$ parties in $G \setminus H$ are byzantine). 

We note that initialization of this tuple is easy: as each honest party $p$ is assumed to have a valid input, it can reliable broadcast the accumulator $h$ associated with that input and all parties in $\mathcal{P}$ can set $G = \{p\}$, with tuple $(h, H) = (h, \{p\})$.

We now present how to merge two groups $G_1$ and $G_2$ with their respective tuples $(h_1, H_1)$ and $(h_2, H_2)$. We note that parts of the procedure, namely sending the value and broadcasting the decision are implementation details, which will be explained in the next section. The only assumption we make about the implementation is that if a group $A$ sends a value to another group $B$, then as long as at least one honest party in $A$ is honest, then every party in $B$ will receive the value.

\begin{dianabox}{\textsc{MergeGroups}}
\algoHead{Merge two disjoint groups $G_1$ and $G_2$}
\begin{algorithmic}
    \State \textbf{Input}: Two groups $G_1$ and $G_2$ with their respective tuples $(h_1, H_1)$ and $(h_2, H_2)$
    \State \textbf{Output}: A valid tuple for $G_1 \cup G_2$

    \If{$H_1 = \emptyset$}
        \State Return $(h_2, H_2)$
    \EndIf
    \State Parties in $H_1$ \textit{send} the value associated with $h_1$ to parties in $H_2$
    \State Parties in $H_2$ \textit{reliable broadcast} $1$ if they received the value associated with $h_1$, $0$ otherwise
    \State Let $K$ be the subset of $H_2$ that broadcasted $0$
    \If{$|K| \geq |H_1|$}
        \State Let $K'$ be a subset of $K$ of size $|H_1|$
        \State Return $(h_2, H_2 \setminus K')$
    \Else
        \State Let $H_1'$ be a subset of $H_1$ of size $|K|$
        \State Return $(h_1, (H_1 \setminus H_1') \cup (H_2 \setminus K))$
    \EndIf
\end{algorithmic}
\end{dianabox}

\begin{lemma}
    When called on two disjoint groups $G_1$ and $G_2$ with their respective tuples $(h_1, H_1)$ and $(h_2, H_2)$ that satisfy the invariant, \textsc{MergeGroups} returns a tuple $(h, H)$ that satisfies the invariant for $G_1 \cup G_2$.
\end{lemma}

\begin{proof}
    Let $M_1 = G_1 \setminus H_1$ and $M_2 = G_2 \setminus H_2$. Under the invariant, both $M_1$ and $M_2$ have a byzantine majority. Because they are disjoint, $M_1 \cup M_2$ also has a byzantine majority. 
    
    We consider the first case in the merge procedure: if $H_1 = \emptyset$, then the tuple $(h_2, H_2)$ is returned. Using the invariant, we already know that honest parties in $H_2$ know the value associated with $h_2$, and the set of remaining parties is exactly $M_1 \cup M_2$, which has a byzantine majority. So the invariant is preserved.

    Otherwise, we remark that either $H_1$ or $K$ is composed only of byzantine parties. Indeed, if $H_1$ is only made of byzantine properties, this remark is true. Otherwise, $H_1$ has at least one honest party, which knows the value to send according to the invariant. Thus, the send will succeed and all honest parties in $H_2$ will receive the value. Hence, $K$, which is made of parties that claimed to not receive the value, is composed only of byzantine parties.

    We remark that in both cases, $(G_1 \cup G_2) \setminus H = M_1 \cup M_2 \cup H_1' \cup K'$, where all these sets are disjoint, $H_1' \subseteq H_1$ and $K' \subseteq K$ such that $|H_1'| = |K'|$. Because $M_1 \cup M_2$ already has a byzantine majority, and one of $H_1'$ or $K'$ is fully composed of byzantine parties, the resulting set will also have a byzantine majority.

    Finally, if $|K| \geq |H_1|$, then the tuple $(h_2, H_2 \setminus K')$ is returned. Because all parties in $H_2$ already claim to know the value with accumulator $h_2$ using the invariant, it is preserved. Otherwise, the tuple $(h_1, (H_1 \setminus H_1') \cup (H_2 \setminus K))$ is returned. Using the invariant, all parties in $H_1$ know the value with accumulator $h_1$, additionally, all parties in $H_2 \setminus K$ broadcasted that they received the value, so they also know it. Thus, the invariant is preserved in both cases.
\end{proof}

\subsection{Implementation}

\paragraph{Sending value} We first explain how to implement the \textit{send} operation. We note that each party in $H_1$ can send the value to all parties in $H_2$. However, if $|H_1| = |H_2| = n/2$, then this would lead to a $\mathcal{O}(n^2 L)$ bit complexity. Instead, we use erasure codes and have each party encode their input into $n$ shares and share it with all parties along with an accumulator witness. This way, when a party in in $H_2$ wants to receive the value, every party can send its share to it and it can reconstruct it. A key point is that if an honest party $p$ is able to reconstruct a valid value, it will immediatly re-encode it and share it again with other parties. This is so that, even if $p$ was only able to get enough shares with the help of byzantine parties, subsequent reconstruction will always succeed, even without their help. We give the implementation of \textsc{Deconstruct} and \textsc{Reconstruct}:
\begin{itemize}
    \item \textsc{Deconstruct}($p, h$): Should only be called if $p$ has the value $v$ with accumulator $h$. $p$ encodes $v$ into $n$ shares along with their witness, then sends each share to the corresponding party. When receiving a share, the party checks that the witness corresponds to the accumulator $h$ and stores it. $p$ sends $\mathcal{O}(\kappa n \log n + L)$ bits, while other parties do not send any bits.
    \item \textsc{Reconstruct}($p, h$): For every party $q$, if $q$ has a share with witness corresponding to $h$, it sends it to $p$. If $p$ receives enough shares (at least $n/2$), it reconstructs the value, checks that its accumulator is $h$, that it is valid and returns it. Otherwise, it returns $\bot$. Each party sends a share, of size $\mathcal{O}(\kappa \log n + L/n)$.
\end{itemize}

\paragraph{Parallel broadcast} We remark that all broadcast operations can be done in parallel. Additionally, if two merges for disjoint groups are done at the same time, their broadcasts can also be done in parallel. We leverage this idea and use our parallel broadcast protocol to perform as many of these broadcasts in parallel. This is needed to make sure the overall round complexity ends up staying linear. Regarding how we merge groups, a first idea would be to merge all groups one by one to a single group. However, this approach cannot be easily parallelized. Another approach is to use a balanced binary tree structure to merge groups, where leafs are the original groups and each internal node represents a merge operation. Although it is more efficient, it would lead to a logarithmic blowup in the bit complexity, as each party would be actively involved in $\mathcal{O}(\log n)$ merges. Instead, we take an approach which is in between these last two: we merge following a tree of height $3$, where each internal node has at most $\lceil n^{1/3} \rceil$ children.

\begin{dianabox}{\textsc{ValidatedAgreement}($\vin$)}
\algoHead{Validated Agreement protocol for a party $p$}
\begin{algorithmic}
    \State Let $h_{in}$ be the accumulator of $\vin$
    \State $(h_p) \gets \textsc{ParallelBA}(\mathcal{P}, h_{in})$
    \For{$p \in \mathcal{P}$, in parallel}
        \State \textsc{Deconstruct}($p, h_p$)
    \EndFor

    \State $G \gets \{ (\{p\}, h_p, \{p\}), p \in \mathcal{P} \}$
    \For{$3$ iterations}
        \State Partition $G$ into sets of size $\lceil n^{1/3} \rceil$
        \State $G' \gets \emptyset$
        \For{each set $S = H_1, \dots, H_k$, in parallel}
            \State $(h, H) \gets H_1$
            \For{$i = 2, \dots, k$}
                \InParallel{for each $p \in H_i$}
                    \State $v \gets \textsc{Reconstruct}(p, h)$
                    \If{$v$ is valid and has accumulator $h$}
                        \State \textsc{Deconstruct}($p, h$)
                    \EndIf
                    \State $b_p \gets 1$ if $v$ is valid and has accumulator $h$, $0$ otherwise
                \EndParallel
                \State Call \textsc{ParallelBA} on $(b_p)_{p \in H_i}$
                \State $(h, H) \gets \textsc{MergeGroups}((h, H), (h_i, H_i))$ using the result of \textsc{ParallelBA}
            \EndFor
            \State $G' \gets G' \cup \{(h, H)\}$
        \EndFor
        \State $G \gets G'$
    \EndFor

    \State Let $h$ be the accumulator of the remaining group in $G$
    \InParallel{for each $p \in \mathcal{P}$}
        \State $v \gets \textsc{Reconstruct}(p, h)$
    \EndParallel
    \State Return $v$
\end{algorithmic}
\end{dianabox}

\begin{lemma}
    The protocol satisfies Validated Agreement.
\end{lemma}

\begin{proof}
    The protocol starts with $n$ groups. In each iteration, it forms sets of size $\lceil n^{1/3} \rceil$ (except the last one) and merges them. As a consequence, after $3$ iterations, there is only $n / n^{1/3^3} = 1$ group left. Consider the last group $G$ with tuple $(h, H)$. By the invariant, $\mathcal{P} \setminus H$ has a byzantine majority. Because $\mathcal{P}$ has a byzantine minority, this implies that $H$ has a byzantine minority and thus has at least one honest party $p$. Using the invariant again, this implies that $h$ is the accumulator of a valid input that $p$ has. We note that every time a party $p$ receives a valid value, it calls \textsc{Deconstruct} on it, which ensures that every honest party will be able to reconstruct the value associated with $h$ at the end of the protocol. Thus, all honest parties will output the same valid value associated with $h$, which satisfies Validated Agreement.
\end{proof}

\begin{lemma}\label{lemma:va-comp}
    \textsc{ValidatedAgreement} has per-party bit complexity:
    $$ \bitc_\fullval(n,L) =  \mathcal{O}(L +  \kappa n b^2 + \kappa n^{1/3} b^3) + \mathcal{O}(n/b + n^{1/3}) \cdot B_{\mvba}(n, b) + \mathcal{O}(n/b) \cdot B_{\mvba}(n, b \kappa)  $$
    And round complexity:
    $$ \roundc_\fullval(n) = \mathcal{O}(n^{1/3}) \cdot (b + \roundc_{\mvba}(n))$$
\end{lemma}

\begin{proof}
    We remark that a call to \textsc{Deconstruct} has a bit complexity of  $\mathcal{O}(\kappa n \log n + L)$ (the sender sends a share and a witness to every other party) to the sender while a call to \textsc{Reconstruct} has a bit complexity of $\mathcal{O}(\kappa n \log n + L/n)$  for every party. In each iteration, which there are $3$, a party is a member of a unique set $H_i$. As a consequence, in each iteration, it will only call \textsc{Reconstruct} at most once and be the sender of at most one \textsc{ParallelBA} operation. Additionally, because we batch all parallel broadcasts calls together for independant merges, each iteration will call at most $k \leq \lceil n^{1/3} \rfloor$ times the \textsc{ParallelBA} procedure. So in total, we have for a given party $p$:
    \begin{itemize}
        \item Initial \textsc{ParallelBA} call: each party broadcasts its accumulator, of size $\kappa$.  Hence, using \cref{lemma:parallel-broad-comp}, we have a total of $(n/b + 1)\left(\mathcal{O}(\kappa b^3) +  \bitc_{\textsc{MVBA}}(n, \kappa b)\right) + \mathcal{O}(\kappa n b^2)$ bits per party.
        \item \textsc{Deconstruct} calls: only the sender sends messages: $1$ at the beginning, at most $1$ in each iteration, for a total of $4 \mathcal{O}(\kappa n \log n + L)$ bits.
        \item \textsc{Reconstruct} calls: at most $n$ in each iteration, for a total of $3n (\mathcal{O}(\kappa \log n + L/n)) = \mathcal{O}(\kappa n \log n + L)$ bits.
        \item \textsc{ParallelBA} calls, ignoring when $p$ is a sender: at most $3n^{1/3}$ calls, all parties are a sender at most once in each iteration. So using \cref{lemma:parallel-broad-comp}, we have a total of $3(n/b + n^{1/3})\left(\mathcal{O}(\kappa b^3) +  \bitc_{\textsc{MVBA}}(n)\right)$ bits per party.
        \item \textsc{ParallelBA} calls, when $p$ is a sender: at most once per iteration, so a total of $3 \cdot \mathcal{O}(\kappa n b^2)$ bits.
    \end{itemize}
    Summing all of these together, we get the expected bit complexity. As for round complexity, the \textsc{Deconstruct} and \textsc{Reconstruct} calls at the beginning and the end take $2$ rounds. Otherwise, in each iteration, we have up to $\lceil n^{1/3} \rceil$ \textsc{ParallelBA} calls which are dependent on each other, so the round complexity becomes: 
    \begin{align*}
        \roundc_\fullval(n) &= 2 + 3\lceil n^{1/3} \rceil \roundc_{ParallelBA}(n) \\
        &=  \mathcal{O}(n^{1/3}) \cdot (b + \roundc_{\mvba}(n)) && \text{Using \cref{lemma:parallel-broad-comp}}
    \end{align*}
\end{proof}
\section{Putting it all together}\label{sec:together}

\subsection{Per-party bit complexity}

We can now combine all the previous protocols together to get a subcubic byzantine agreement protocol. We first start by simplifying the complexity of the partial validated agreement protocol. Using \cref{lemma:pva-comp}, we have:
\begin{align*}
    \bitc_\pval(n,L) = \mathcal{O}(L \log n + \kappa n b^2 + \kappa b^3 \log n) + n/b \cdot  \bitc_\mvba(n, b) + \mathcal{O}(\log n) \cdot \bitc_{\mvba}(n, L)  \\ 
        + \sum_{i = 0}^{\lceil \log_{3/2} n \rceil}  \bitc_\fullval(n \cdot (2/3)^i, L)
\end{align*}
Using \cref{lemma:va-comp}, we have:
\begin{align*}
    \bitc_\fullval(n,L) =  \mathcal{O}(L +  \kappa n b^2 + \kappa n^{1/3} b^3) + \mathcal{O}(n/b + n^{1/3}) \cdot B_{\mvba}(n, b) + \mathcal{O}(n/b) \cdot B_{\mvba}(n, b \kappa)  
\end{align*}
Using the fact that $\sum_{i = 0}^{\lceil \log_{3/2} n \rceil} (n \cdot (\alpha)^i) = \mathcal{O}(n)$ for any $0 < \alpha < 1$, we have:
\begin{align*}
    \sum_{i = 0}^{\lceil \log_{3/2} n \rceil}  \bitc_\fullval(n \cdot (2/3)^i, L) &= \mathcal{O}(L \log n +  \kappa n b^2 + \kappa n^{1/3} b^3) \\& + \mathcal{O}(n/b + n^{1/3}) \cdot B_{\mvba}(n, b) + \mathcal{O}(n/b) \cdot B_{\mvba}(n, b \kappa)  
\end{align*} 

Using these two equations together, we get:
\begin{align*}
    \bitc_\pval(n,L) &= \mathcal{O}(L \log n + \kappa n b^2 + \kappa n^{1/3} b^3) + \mathcal{O}(n/b + n^{1/3}) \cdot  \bitc_\mvba(n, b) \\& + \mathcal{O}(n/b) \cdot  \bitc_\mvba(n, \kappa b) + \mathcal{O}(\log n) \cdot \bitc_{\mvba}(n, L) \\
    & = \mathcal{O}(L \log n + \kappa n b^2 + \kappa n^{1/3} b^3 + \kappa n^2 \log n / b + \kappa n^{4/3} \log n) \\ &+ \mathcal{O}(n/b + n^{1/3})\cdot \kappa \cdot \bitc_{\ba}(n) && \text{Using \cref{lemma:mvba-comp}} 
\end{align*}

We now look at the complexity of our weak agreement protocol:
\begin{align*}
    \bitc_{\wba}(n) &= \bitc_\proofcrea(n) + \bitc_\valdis(n, \kappa n) && \text{Using \cref{obs:weak-comp}} \\
    & = \mathcal{O}(n(\log^2 n + \kappa)) + \bitc_\valdis(n, \kappa n) && \text{Using \cref{theo:proof-crea-comp}} \\
    & = \mathcal{O}(n(\log^2 n + \kappa)) + \mathcal{O}(n m \kappa ) + \bitc_{\textsc{PVA}}(\lceil n/m \rceil, \kappa n) && \text{Using \cref{theo:vd-comp}} \\
    & = \mathcal{O}(n m \kappa + \kappa n b^2 / m + \kappa n^{1/3} b^3 / m^{1/3} + \kappa n^2 \log n / (b m^2) + \kappa n^{4/3} \log n / m^{4/3}) \\ &+ \mathcal{O}(n/(m b) + (n/m)^{1/3}) \cdot \kappa \cdot \bitc_{\ba}(\lceil n/m \rceil)  
\end{align*}

Finally, we can consider the complexity of the binary agreement protocol, for $n$ big enough, we have using \cref{lemma:ba-comp,lemma:graded-comp}:
\begin{align*}
    \bitc_{\ba}(n) \leq 2 \bitc_{\ba}(\lceil n/2 \rceil) + 4 \bitc_{\wba}(n)
\end{align*}
So in total, we have: 
\begin{align*}
    \bitc_{\ba}(n)& \leq 2 \bitc_{\ba}(\lceil n/2 \rceil) +  \mathcal{O}(n m \kappa + \kappa n b^2 / m + \kappa n^{1/3} b^3 / m^{1/3} + \kappa n^2 \log n / (b m^2) + \kappa n^{4/3} \log n / m^{4/3}) \\ &+ \mathcal{O}(n/(m b) + (n/m)^{1/3}) \cdot \kappa \cdot \bitc_{\ba}(\lceil n/m \rceil)  
\end{align*}

\begin{theorem}\label{theo:ba-final-comp}
    For $n$ large enough, by choosing $m = b = \Theta((\kappa n)^{(\sqrt{13} - 3)/2}) \approx \Theta((\kappa n)^{0.3077})$, we obtain
    \begin{align*}
        \bitc_{\ba}(n) = \mathcal{O}((\kappa n)^{(\sqrt{13} - 1)/2})
    \end{align*}
\end{theorem}

\begin{proof}
    We note that the depency on $\kappa$, which cannot be considered as $\mathcal{O}(1)$ in $\bitc_{\ba}(n)$'s relation prevents us from using the Akra-Bazzi method. Our proof relies on the fact that $\kappa = \text{polylog} \ n$. This implies that there are some constants $C,K > 0$ such that $\kappa \leq K \log^C n$. One thing to note is that although this inequality should be true for the initial number of parties $n$, it may not hold in the number of parties $n'$ considered in recursive calls, as that would imply that $\kappa = \mathcal{O}(1)$. Let us define $p = (\kappa n)^{(\sqrt{13} - 1)/2} \approx 1.3077$. To give some informal insight regarding the parameters used, we will use $m(n) = b(n) \approx (\kappa n)^{p-1} \approx (\kappa n)^{0.3077}$.

    Let $N > 1$ large enough such that $\kappa \leq K \log^C N$. We want to show that there exists constant $A$ depending only on $K$ and $C$ such that $\bitc_{\ba}(n) \leq A (\kappa n)^p$.

    To do so, we prove by induction that for all $1 < n \leq N$, we have:
    \begin{align*}
        \bitc_{\ba}(n) \leq A (\kappa n)^p \left( \frac{\log(N)}{\log(n)} \right)^D
    \end{align*}
    With $D$ a constant to be determined later. We note that if this property holds for $n = N$, then we have the expected result.

    We now give the explicit values of $m$ and $n$ used to obtain this bound:
    \begin{align*}
        (m,b) = \begin{cases}
        (n, 1) & \text{If $n \leq B \kappa^4$} \\
        (c(\kappa n)^{p-1}, c(\kappa n)^{p-1}) & \text{Otherwise}
        \end{cases}
    \end{align*}
    Where $B$ and $c$ are large enough constants to be determined later. 

    Let us define $f(n) = A (\kappa n)^p \left( \frac{\log(N)}{\log(n)} \right)^D$. We consider the first case, where $1 < n \leq B \kappa^4$. In this case, we use $m = n$ and $b = 1$. We note when looking at the validated dispersal protocol that this effectively amounts to parties sending their proof to all other parties if they have one. In fact, looking at the recurrence we have on $\bitc_{\ba}(n)$, we obtain:
    \begin{align*}
        \bitc_{\ba}(n) &\leq 2 \bitc_{\ba}(\lceil n/2 \rceil) +  \mathcal{O}(\kappa n^2 + n \cdot \kappa \cdot \bitc_{\ba}(1)) \\
            & \leq 2 \bitc_{\ba}(\lceil n/2 \rceil) +  \mathcal{O}(\kappa n^2 )
    \end{align*} 
    Because byzantine agreement on a single party needs no communication ($\bitc_{\ba}(1) = 0$). A simple induction argument shows that $\bitc_{\ba}(n) = \sum_{i = 0}^{\log n} \mathcal{O}(\kappa) \cdot 2^i (n/2^i)^2 = \mathcal{O}(\kappa n^2)$. Hence, we can find $\alpha > 0$ such that:
    \begin{align*}
        \bitc_{\ba}(n) &\leq \alpha \cdot \kappa n^2 \\
        & \leq \alpha \cdot (\kappa n)^p n  && \text{Because $p \geq 1$}\\
        & \leq \alpha \cdot (\kappa n)^p B \kappa^4 && \text{Because $n \leq B \kappa^4$} \\
        & \leq \alpha \cdot B \cdot (\kappa n)^p K^4 \log^{4C} N 
    \end{align*}
    We remark that we assume $\kappa \leq K \log^C N$ and $n \leq B \kappa^4$ so $\log n = o_N(\log^\beta N)$ for any $\beta > 0$. Therefore, for $N$ large enough, if $D > 4C$, we can find $B$ large enough such that:
    \begin{align*}
        \bitc_{\ba}(n) &\leq A (\kappa n)^p \left( \frac{\log(N)}{\log(n)} \right)^D = f(n)
    \end{align*} 

    We can now consider the second case, where $n > B \kappa^4$. Fix $B \kappa^4 < n \leq N$ and assume the invariant already holds for all $1 < n' < n$. We have $m = b = c (\kappa n)^{p-1} \approx \Theta(\kappa n)^{0.3077}$. As a consequence, we can remove out lower-order terms in the recurrence of $\bitc_{\ba}(n)$ and simplify it:  
    \begin{align*}
        \bitc_{\ba}(n) &\leq 2 \bitc_{\ba}(\lceil n/2 \rceil) + \mathcal{O}(c(\kappa n)^p + c^{-2}(\kappa n)^{3 - 2p} \cdot \bitc_{\ba}(\lceil n/m \rceil)) \\
        & = 2 \bitc_{\ba}(\lceil n/2 \rceil) + C_1 \cdot c (\kappa n)^p + C_2 \cdot c^{-2} (\kappa n)^{3 - 2p} \cdot \bitc_{\ba}(\lceil n/m \rceil)
    \end{align*}
    For some constant $C_1$ and $C_2$. First, using the induction hypothesis:
    \begin{align*}
        2 \bitc_{\ba}(\lceil n/2 \rceil) &\leq 2 A (\kappa n/2)^p \left( \frac{\log(N)}{\log(\lceil n/2 \rceil)} \right)^D \\ 
        & \leq 2^{1-p}\cdot \left( \frac{\log(\lceil n/2 \rceil)}{\log(n)} \right)^D f(n)
    \end{align*}
    We note that $\lim_{n \to \infty}  2^{1-p}\cdot \left( \frac{\log(\lceil n/2 \rceil)}{\log(n)} \right)^D = 2^{1-p} < 1$. Let $h = (2^{1-p} + 1)/2 < 1$. For $n$ large enough (which we can ensure by taking $B$ large enough), we therefore have $2 \bitc_{\ba}(\lceil n/2 \rceil) \leq h f(n)$. We consider the second recursive term, using the induction hypothesis again and the fact that $\lceil n/m \rceil \leq 2 n/m$:
    \begin{align*}
        C_2 \cdot c^{-2}(\kappa n)^{3 - 2p} \cdot \bitc_{\ba}(\lceil n/m \rceil) &\leq C_2(\kappa n)^{3 - 2p} \cdot A (\kappa \lceil n/m \rceil)^p \left( \frac{\log(N)}{\log(\lceil n/m \rceil)} \right)^D \\
        & \leq C_2 \cdot c^{-2}(\kappa n)^{3 - 2p} (2/m)^p  \left( \frac{\log n}{\log(\lceil n/m \rceil)} \right)^D f(n) \\
        & \leq C_2 \cdot c^{-2 - p}(\kappa n)^{3 - p - p^2} 2^p \left( \frac{\log n}{\log c + (2 -p) \log(n)} \right)^D f(n) \\
    \end{align*}
    We remark that $p = (\sqrt{13} - 1)/2$ is a solution of $3 - x - x^2 = 0$. Because $\lim_{n \to \infty}  \frac{\log n}{\log c + (2 -p) \log(n)}  = 1/(2 - p) \approx 1.43$, for $n$ large enough (which we can ensure again by taking $B$ large enough), we have:
    \begin{align*}
        C_2 \cdot c^{-2}(\kappa n)^{3 - 2p} \cdot \bitc_{\ba}(\lceil n/m \rceil) &\leq C_2 \cdot c^{-2 - p} 2^p 1.5^D \cdot f(n) \\
        &\leq (1-h)/2 f(n) && \text{By taking $c$ large enough}
    \end{align*}
    Finally, by taking $A$ large enough such that $C_1 \cdot c \leq A \cdot (1-h)/2$, we have:
    \begin{align*}
        \bitc_{\ba}(n) &\leq h f(n) + (1-h)/2 f(n) + (1-h)/2 f(n) = f(n)
    \end{align*}
    So the invariant is preserved. This concludes the proof.
\end{proof}

\subsection{Round complexity}

In the previous section, we described how to choose the parameters $m$ and $b$ to obtain the desired complexity. We now show that given these parameters, our protocol has an optimal linear round complexity.

Using \cref{lemma:pva-comp}, we have:
\begin{align*}
    \roundc_\pval(n) = \mathcal{O}(\log n) (b + \roundc_\mvba(n)) + \sum_{i = 0}^{\lceil \log_{3/2} n \rceil}  \roundc_\fullval(n \cdot (2/3)^i)
\end{align*}
Using \cref{lemma:va-comp}, we have:
\begin{align*}
    \roundc_\fullval(n) = \mathcal{O}(n^{1/3}) \cdot (b + \roundc_{\mvba}(n))
\end{align*}
Using the fact that $\sum_{i = 0}^{\lceil \log_{3/2} n \rceil} (n \cdot ((2/3)^{1/3})^i) = \mathcal{O}(n)$ , we have:
\begin{align*}
    \sum_{i = 0}^{\lceil \log_{3/2} n \rceil} \roundc_\fullval(n \cdot (2/3)^i) &=  \mathcal{O}(n^{1/3}) \cdot (b + \roundc_{\mvba}(n))
\end{align*} 

Using these two equations together, we get:
\begin{align*}
    \bitc_\pval(n,L) &=  \mathcal{O}(n^{1/3}) \cdot (b + \roundc_{\mvba}(n))
\end{align*}

We now look at the complexity of our weak agreement protocol:
\begin{align*}
    \roundc_{\wba}(n) &= \roundc_\proofcrea(n) + \roundc_\valdis(n, \kappa n) && \text{Using \cref{obs:weak-comp}} \\
    & = \mathcal{O}(\log n) + \roundc_\valdis(n) && \text{Using \cref{theo:proof-crea-comp}} \\
    & = \mathcal{O}(\log n) + \mathcal{O}(1) + \roundc_{\textsc{PVA}}(\lceil n/m \rceil) && \text{Using \cref{theo:vd-comp}} \\
    & = \mathcal{O}((n/m)^{1/3})(b + \roundc_{\mvba}(\lceil n/m \rceil)) \\
    & =  \mathcal{O}((n/m)^{1/3})(b + \roundc_{\ba}(\lceil n/m \rceil)) && \text{Using \cref{lemma:mvba-comp}}
\end{align*}

Finally, we can consider the complexity of the binary agreement protocol, for $n$ big enough, we have using \cref{lemma:ba-comp,lemma:graded-comp}:
\begin{align*}
    \roundc_{\ba}(n) \leq 2 \roundc_{\ba}(\lceil n/2 \rceil) + 4 \roundc_{\wba}(n)
\end{align*}
So in total, we have: 
\begin{align*}
    \roundc_{\ba}(n)& \leq 2 \roundc_{\ba}(\lceil n/2 \rceil) + \mathcal{O}((n/m)^{1/3})(b + \roundc_{\ba}(\lceil n/m \rceil))
\end{align*}

\begin{theorem}\label{theo:ba-full-round}
    Using the same parameters for $b$ and $m$ as in \cref{theo:ba-final-comp}, $\roundc_{\ba}(n) = \mathcal{O}(n)$.
\end{theorem}

\begin{proof}
    As a reminder, in \cref{theo:ba-final-comp}, we use 
    \begin{align*}
        (m,b) = \begin{cases}
        (n, 1) & \text{If $n \leq B \kappa^4$} \\
        (c(\kappa n)^{p-1}, c(\kappa n)^{p-1}) & \text{Otherwise}
        \end{cases}
    \end{align*}
    For some constants $B$ and $c$ large enough and $p = (\sqrt{13} - 1)/2$. We show by induction that for all $n \geq 1$:
    \begin{align*}
        \roundc_{\ba}(n) \leq A (n - n^{0.95}/2)
    \end{align*}
    For some constant $A$ large enough. When $n = 1$, the protocol returns immediately its input, so $\roundc_{\ba}(1) \leq 1$. 
    Assume $1 < n \leq B \kappa^4$. In this case, we have $m = n$ and $b = 1$, so:
    \begin{align*}
        \roundc_{\ba}(n)& \leq 2 \roundc_{\ba}(\lceil n/2 \rceil) + \mathcal{O}(1) \cdot (1 + \roundc_{\ba}(1)) = 2 \roundc_{\ba}(\lceil n/2 \rceil) + \mathcal{O}(1) 
    \end{align*}
    A simple induction argument shows that $\roundc_{\ba}(n) \leq \sum_{i = 0}^{\log n} \mathcal{O}(1) \cdot 2^i = \mathcal{O}(n)$, so with $A$ large enough, we have $\roundc_{\ba}(n) \leq A n/2 \leq A (n - n^{0.95}/2)$.
    We now consider $n > B \kappa^4$. In this case, we have $m = b = c (\kappa n)^{p-1}$. Because $\kappa^4 < n$, we have $\kappa = \mathcal{O}(n^{1/4})$. Additionally, $b = m = \Theta((\kappa n)^{p-1}) = \Omega(n^0.3)$ and $m = b = \mathcal{O}(n^{0.31})$.  Therefore:
    \begin{align*}
        \mathcal{O}((n/m)^{1/3})(b + \roundc_{\ba}(\lceil n/m \rceil)) &= \mathcal{O}((n^{0.7})^{1/3})((\kappa n)^{0.31} + \roundc_{\ba}(\lceil n/m \rceil)) \\
        & \leq \mathcal{O}(n^{0.24})((n^{1.25})^{0.31} + A(2n^0.7 - n^{0.7 * 0.95}/2)) \\
        & = \mathcal{O}(n^{0.94}) 
    \end{align*}
    For $n$ large enough, which can be ensured by taking $B$ large enough. Moreover, using the induction hypothesis, we have:
    \begin{align*}
        2 \roundc_{\ba}(\lceil n/2 \rceil) &\leq 2 A (\lceil n/2 \rceil - \lceil n/2 \rceil^{0.95}/2) \\
        & \leq A(n - n^{0.95}/2) + 2A - (\lceil n/2 \rceil^{0.95} - n^{0.95}/2) \\
        & \leq A(n - n^{0.95}/2) - \Omega(n^{0.95})
    \end{align*}
    Therefore, putting it together, we have:
    \begin{align*}
        \roundc_{\ba}(n) &\leq A(n - n^{0.95}/2) - \Omega(n^{0.95}) + \mathcal{O}(n^{0.94})  \\
        & \leq A(n - n^{0.95}/2)
    \end{align*}
    For $n$ large enough, which can be ensured by taking $B$ large enough. The induction is preserved, so this concludes the proof: we have $\roundc_{\ba}(n) = \mathcal{O}(n)$. So the protocol is linear.
\end{proof}

\subsection*{Final Result}

\byzantineagreement*

\begin{proof}
    Using \cref{theo:ba-full-round}, the round complexity of the protocol is linear and with \cref{theo:ba-final-comp}, the per-party bit complexity is $\mathcal{O}(\kappa^{(\sqrt{13} - 1)/2}n^{(\sqrt{13} + 1)/2})$. Multiplying this by $n$, we get the expected upper bound on the total bit complexity.
\end{proof}

\section{Message and signature tags}\label{sec:tag}

We note that the protocols we have described so far have a main weakness: we expect some of them to run in parallel, but messages do not get sufficient information to know which instance of the recipient party it targets. Additionally, our protocols as they are written are weak to replay attacks: a byzantine could take a vote from an earlier or parallel vote and use it in a later vote for a different value. To fix this, we explain how to add tags and prefixes to messages and signatures to ensure that they are unique to a given instance of a protocol.

First, we remark that given an instance of byzantine agreement of size $n$, we never have instances of size $n/2$ running concurrently. Only instances of size $n/m$ will. Moreover, in each step, we only run at most a linear number of instances in parallel, and there is a bonded number of steps in a given recursion depth. As a consequence, a given instance executing at a given time and recursion depth with $n$ parties can be uniquely identified with a $\mathcal{O}(\log n)$-bit tag. As a consequence, the total tag size $T(n)$ needed for any instance at a given time satisfies:
\begin{align*}
    T(n) = \mathcal{O}(\log n) + T(n/m)
\end{align*}
Because, as seen in \cref{theo:ba-final-comp}, we use $m = \Omega(n^\alpha)$ for some $\alpha > 0$, an immediate induction shows that $T(n) = \mathcal{O}(\log n)$. Therefore, a $\mathcal{O}(\log n)$-bit tag added as a prefix to each message is sufficient to ensure that a party can identify which instance of the protocol a message targets. We remark that all messages exchanged in our protocols already have size $\Omega(\log n)$, so this does not change the bit complexity of our protocols.

We now consider signatures and replay attacks. Adding the previous tag as a prefix to signatures is enough to prevent replay attacks across concurrent instances. However, they are not enough for attacks executed using a signature from an earlier instance in a later instance. A simple workaround is to add the current round number as an additional prefix to each signature, and send it along the signature. Because the protocol has a linear number of rounds, this only adds an additional $\mathcal{O}(\log n) = \mathcal{O}(\kappa)$ bit to each signature. If the round number of a signature does no match what is expected of the current instance, the whole message is discarded.

\section{Conclusion}

In this paper, we present the first subcubic deterministic byzantine agreement protocol in the synchronous setting with optimal resiliency $t < n/2$. To do so, we introduce multiple new approaches, including better recursive techniques, use of fault-tolerant expanders and dispersers as well as fair cake-cutting techniques. We remark that our protocol has an unusual irrational exponent, being $(\sqrt{13} + 1)/2$. 

There are still open questions regarding this problem. The first one would be to close the gap, which is still of a factor $n^{0.303}$ between the quadratic lower bound of Dolev and Reischuk \cite{PODC:DolRei82} and our upper bound. We remark that our proof gathering protocol is already near-optimal. As such, any improvement to validated dispersal, which we believe could be simplified would yield a more efficient algorithm.


\appendix
\newpage

\section{Expander and Disperser Graphs}

A graph is said to be explicit if it can be constructed deterministically in polynomial time. 

\begin{theorem}\label{theo:big-disperser}
    For any constant $0 < \alpha < 1$, $1 \leq m \leq n$, there exists an explicit directed graph $G = (V, E)$ on $n$ vertices where each vertex has out-degree m and such that $G$ is a $(\alpha n, n - 4 \alpha^{-1} n/m)$-disperser.
\end{theorem}

\begin{proof}
    Let $1 \leq m \leq n$, we explain how to build explicitly such a graph and how to get $G$ from it.

    We consider a regular Ramanujan bipartite graph $H = L \cup R$ on $2n$ nodes ($n$ on each side) with degree $d = m$. Cohen \cite{FOCS:Cohen16b} explains how to build such a graph in polynomial time. We want to show that this graph is a $(\alpha n, n - 4 \alpha^{-1} n/m)$ bipartite disperser. To do so, consider a set $S$ of $\alpha n$ vertices in $L$, let $N(S)$ be the neighbors of $S$ in $R$. We consider $T = R \setminus N(S)$, there is no edge between $S$ and $T$. We want to prove that $|T| \leq 4 \alpha^{-1} n/m$.
    
    We remark that by definition of a bipartite Ramanujan graph, the non-trivial eigenvalues of $H$ have absolute value at most $2\sqrt{d -1}$. We can therefore use the bipartite expander mixing lemma (a proof of it can be found in \cite[Lemma 4.1]{bipartite_mixing_lemma}):

    \begin{align*}
        \left| e(S, T) - d \frac{|S| |T|}{n} \right| \leq 2 \sqrt{d-1} \cdot \sqrt{|S| |T|}
    \end{align*}
    Where $e(S,T)$ is the number of edges between $S$ and $T$, which by our choice of $T$ is $0$. Using $d = n^\alpha$:
    \begin{align*}
        d \frac{|S| |T|}{n} &\leq 2 \sqrt{d}  \cdot \sqrt{|S| |T|} \\
        \sqrt{d |S| |T|} &\leq 2n \\
        d |S| |T| &\leq 4n^2 \\
        m(\alpha n) |T| &\leq 4n^2 \\
        |T| &\leq 4 \alpha^{-1} n/m
    \end{align*}
    Therefore, $N(S) \geq n - 4 \alpha^{-1} n/m$, So $G_1$ is indeed a bipartite disperser.

    To build $G = (V,E)$ we then identify $G_1$ with the incidence graph of $G$, i.e. we identify $V$ with both $L$ and $R$, and we add an edge $(u,v)$ in $G$ if there exists an edge from $u$ in $L$ to $v$ in $R$. By construction, $G$ is a directed graph with out-degree $m$ and is a $(\alpha n, n - 4 \alpha^{-1} n/m)$-disperser.
\end{proof}

Using the previous theorem, we obtain a biparte version where both sides are not equal:

\begin{theorem}\label{theo:small-bipartite}
    For any $1 \leq m \leq n$, there exists an explicit bipartite graph $G = (L \cup R, E)$ with $|L| = n$, $|R| = m$ and constant left degree such that $G$ is a $(n/3, 5m/6)$ bipartite disperser.
\end{theorem}

\begin{proof}
    As seen in the proof of the previous theorem, we can build an explicit bipartite graph $G_1 = (L_1 \cup R_1, E_1)$ with $|L_1| = |R_1| = n$ such that $G_1$ is a $(n/3, n - n/18)$ bipartite disperser by using $\alpha = 1/3$ and degree $d = 4 * 3 * 18 = \mathcal{O}(1)$. 

    We now partition $R_1$ into $m$ groups, each of size $\lfloor n/m \rfloor$ or $\lceil n/m \rceil$. Because $m \leq n$, $\lfloor n/m \rfloor \geq \lceil n/m \rceil /2$. We then build $G$ by identifying $L$ with $L_1$ and $R$ with the $m$ groups of $R_1$ and defining the edges such that for each edge $(u, v) \in E_1$, we add an edge $(u, v') \in E$ where $v'$ is the group to which $v$ belongs. By construction, $G$ has left degree $d = 4 * 3 *18$ which is constant.

    We note that $G$ can be a multigraph, but each edge of $G$ can be mapped to an edge in $G_1$. Because $G_1$ is $d$-regular and $\lfloor n/m \rfloor \geq \lceil n/m \rceil /2$, This implies that vertices in $R$ are composed of at least $n/(3m)$ vertices from $R_1$. Consider a set $S \subseteq L$ of size $|S| \geq n/10$, let $Q = R \setminus N_G(S)$. We want to show that $|Q| \leq m/6$. We note that $N_{G_1}(S) \geq n-n/18$, which would imply that $Q$ is composed of at most $n/18$ parties in $R_1$. Because each party in $R$ is composed of at least $n/(3m)$ vertices from $R_1$, and there cannot be an edge from $S$ to $Q$, we have $|Q| \leq (n/18) / (n/(3m)) = m/6$. Therefore, $G$ is a $(n/10, 5m/6)$ bipartite disperser.
\end{proof}

For the following theorem, we first show an intermediate lemma:

\begin{lemma}\label{lemma:app-inter-rama}
    Let $G = (L \cup R, E)$ be a $D$-regular Ramanujan bipartite graph, with each side having $n$ nodes. Let $S \subset L$ and $H = \{ v \in R, N(v) \subseteq H\}$ (i.e $H$ is the set of nodes whose edges all come from $S$). Then:
    \begin{align*}
        |H| \leq \frac{4|S|}{D \cdot (1- |S|/n)^2}
    \end{align*}
\end{lemma}

\begin{proof}
    By definition, we have $e(S, H) = |H| \cdot D$. We can therefore use the bipartite expander lemma on $S$ and $H$ (\cite[Lemma 4.1]{bipartite_mixing_lemma}):
    \begin{align*}
        \left| e(S, H) - D \cdot \frac{|S||H|}{n} \right| &\leq 2\sqrt{D-1} \sqrt{|S||H|} \\
        D \cdot |H| \left( 1 -  |S|/n\right) &\leq 2 \sqrt{D |S||H|} \\
        D \cdot |H| \left( 1 -  |S|/n\right)^2 &\leq 4 |S| \\
        |H| &\leq \frac{4|S|}{D \cdot (1- |S|/n)^2}
    \end{align*}
\end{proof}

\expander*

\begin{proof}
    Let $n > 0$, the graph we obtain uses the construction from \cite{PODC:Upfal92}, which is essentially a weaker version of this theorem with fixed $\alpha = 1/72$ and $\beta = 6$, as a basis. Using \cite{PODC:Upfal92}, there exists $D_1 > 0$ independent of $n$ and an explicit $D_1$-regular graph $G_1 = (V_1, E_1)$ on $n$ vertices such that for any $C \subset V_1$ satisfying $|C| \leq n/72$, one can find $S \subset V_1$ such that $S \cap C = \emptyset$, $|S| \geq n - 6 |S|$ and the induced subgraph $G_1[S]$ is connected and of diameter $\mathcal{O}(\log n)$.

    We now consider $D_2 = D_2(\alpha, \beta)$ which we will explicit later in this proof. We consider the $D_2$ regular bipartite Ramanujan graph $G_2 = (L \cup R, E_2)$ with $n$ vertices on each side, such a graph can be built in polynomial time \cite{FOCS:Cohen16b}. We now explain how to construct the resulting graph $G = (V,E)$. We identify $V$ with $L$ and $V_1$ with $R$. Then:
    \begin{align*}
        E = \{(u,v), \exists (u', v'), (u, u') \in E_2, (v,v') \in E_2, (u', v') \}
    \end{align*}
    What we essentially do is layer $G_2$ on top of $G_1$: $G_2$ maps vertices to groups (a vertex belongs to $D_2$ groups) two vertices are linked in $G$ if they each belong to a group which are connected in $G_1$. A consequence is that vertices have degree at most $D_1 \cdot D_2$ which is independent of $n$. 

    We now want to prove that $G$ satisfies the properties of the theorem with a proper choice of $D_2$. Let $C \subset V$ such that $|C| \leq \alpha \cdot n$. Let $H = \{ v \in R, N_{G_2}(v) \subseteq C\}$. We want to show that $|H| \leq n/72$ and that $|H| = \mathcal{O}(1)\cdot |C|$. Using \cref{lemma:app-inter-rama}, we have $|H| \leq \frac{4|C|}{D \cdot (1- |C|/n)^2}$. We remark that $|C| \leq \alpha \cdot n$ so $|H| \leq \frac{4|C|}{D \cdot (1- \alpha)^2} \leq \frac{4n}{D \cdot (1- \alpha)^2}$. 

    Therefore,  by choosing $D_2 \geq 288 / (1 - \alpha)^2$, we have $|H| \leq n/72$. As a consequence, by construction of $G_1$, one can find $K \subseteq V_1$ such that $K \cap H = \emptyset$, $|K| \geq n - 6 |H| \geq 11n/12$ and $G_1[K]$ is connected with diameter $\mathcal{O}(\log n)$.

    Let $S = N_{G_2}(K) \setminus C$. We want to show that $G[S]$ is connected, has diameter $\mathcal{O}(\log n)$ and $|S| \geq n - \beta |C|$. To do so, consider $v_1, v_2 \in S$, by definition, we can find $w_1, w_2 \in K$ such that $(v_1, w_1), (v_2, w_2) \in E_2$. Because $G_1[K]$ has diameter $\mathcal{O}(\log n)$, one can find $w_1= u_1, u_2, \ldots, u_k = w_2$ with $k = \mathcal{O}(\log n)$ such that $u_1 \rightarrow u_2 \rightarrow \ldots \rightarrow u_k$ forms a path in $G_2$. By definition of $K$, for each $u_i$, one can find $p_i \in V \setminus C$ such that $(p_i, u_i) \in E_1$, so $p_i \in S$ (otherwise we would have $u_i \in H$). We remark that $v_1 \rightarrow p_2 \rightarrow \ldots \rightarrow p_{k-1} \rightarrow v_2$ is by construction a path of length $k = \mathcal{O}(\log n)$ between $v_1$ and $v_2$ contains within $S$.

    It remains to show that $|S| \geq n - \beta |C|$ to complete the proof. Because $S = N_{G_2}(K) \setminus C$, it is enough to show that $|N_{G_2}(K)| \geq n - (\beta -1)|C|$ (which is possible as $\beta > 1$). Let $P = \{u \in L, N(u) \cap K = \emptyset\}$. We remark that by definition, $N_{G_2}(K)$ and $P$ form a partition of $L$, so $|N_{G_2}(K)| + |P| = n$ and thus it is enough to prove that $|P| \leq (\beta - 1)\cdot |C|$. We remark that $P$ can be rewritten as $P = \{u \in L, N(u) \subseteq R \setminus K\}$, so we can use \cref{lemma:app-inter-rama} (both sides of a Ramanujan graph have the same properties):
    \begin{align*}
        |P| &\leq \frac{4 |R \setminus K |}{D_2(1 - |R\setminus K|/n)^2} \\
        &\leq \frac{4 (n - |K|)}{D_2(1-(n-|K|)/n)^2} \\
        &\leq \frac{4\cdot 6 |H|}{D_2\cdot (1/12)^2} && \text{Because $|K| \geq n - 6 |H| \geq 11n/12$} \\
        &\leq 3456 \cdot \frac{4 |C|}{D_2^2(1-\alpha)^2}
    \end{align*}
    Thus, by taking $D_2 \geq \frac{\sqrt{13824}}{(1-\alpha)^2 (\beta-1)}$, we get $|P| \leq (\beta-1)n$ and in turn $|S| \geq n - \beta |C|$, so $S$ satisfies the conditions.

    Finally, by taking $D_2 = D_2(\alpha, \beta) = \max\left(\frac{288} { (1 - \alpha)^2}, \frac{\sqrt{13824}}{(1-\alpha)^2 (\beta-1)} \right)$, the assumptions above are held.
\end{proof}
\section{Missing protocols}

We provide here the missing protocols used in our paper. The reason they are only presented here is that they are common knowledge or straightforward implementation of existing work.

\subsection{Byzantine Agreement}\label{subsec:ba-protocol}

We present here the framework given by Momose and Ren \cite{momoseren} to reduce byzantine agreement to weak agreement. We note that their protocol only reduces byzantine agreement to graded agreement, which we give a definition below. We also give a simple reduction from graded agreement to weak agreement, which together show the overall reduction.

\begin{definition}
    A protocol where each party is given an input $v_{in} \in \{0,1\}$ and decides an output $v_{out} \in \{0,1\}$ and a grade $g \in \{0,1\}$ satisfies graded agreement if:
    \begin{itemize}
        \item Termination: All honest parties decide
        \item Validity: If all honest parties have the same input $v$, then only $v$ can be decided with grade $1$
        \item Graded Agreement: If an honest party decides $v$ with grade $1$, then all honest parties decide $v$.
    \end{itemize}
\end{definition}

\begin{dianabox}{\textsc{ByzantineAgreement}($\vin$)}
\algoHead{Protocol for byzantine agreement for a party $p$ with input $\vin$}
\begin{algorithmic}
    \State If $|\mathcal{P}| < M$ for some constant $M$, use an inefficient cubic byzantine agreement protocol instead.

    \State Partition $\mathcal{P}$ into two halves $P_1$ and $P_2$.
    \State $(v,g) \gets \textsc{GradedAgreement}_\mathcal{P}(\vin)$
    \If{$p \in P_1$}
        \State $o \gets \textsc{ByzantineAgreement}_{P_1}(v)$
        \State Send $o$ to all parties
    \EndIf
    \If{$g = 0$ and received the same $o$ from a majority of $P_1$}
        \State $v \gets o$
    \EndIf

    \Statex
    \State $(v,g) \gets \textsc{GradedAgreement}_\mathcal{P}(v)$
    \If{$p \in P_2$}
        \State $o \gets \textsc{ByzantineAgreement}_{P_2}(v)$
        \State Send $o$ to all parties
    \EndIf
    \If{$g = 0$ and received the same $o$ from a majority of $P_2$}
        \State $v \gets o$
    \EndIf
    \State Return $v$
\end{algorithmic}
\end{dianabox}

\begin{theorem}{\cite[Lemma 9]{momoseren}}
    \textsc{ByzantineAgreement} Satisfies byzantine agreement.
\end{theorem}

\begin{lemma}\label{lemma:ba-comp}
    The protocol is byzantine robust, has round complexity $\roundc_{\ba}(n) = \roundc_{\ba}(\lfloor n/2 \rfloor) + \roundc_{\ba}(\lceil n/2 \rceil) + 2 \roundc_\textsc{GBA}(n) + \mathcal{O}(1)$ and per-party bit complexity $\bitc_{\ba}(n) = \bitc_{\ba}(\lfloor n/2 \rfloor) + \bitc_{\ba}(\lceil n/2 \rceil) + 2 \bitc_\textsc{GBA}(n) + \mathcal{O}(n \kappa)$.
\end{lemma}

\subsection{Graded agreement}\label{subsec:gba-protocol}

\begin{dianabox}{\textsc{GradedAgreement}($\vin$)}
\algoHead{Protocol for graded agreement for a party $p$ with input $\vin$}
\begin{algorithmic}
    \State $v \gets \textsc{WeakAgreement}(\vin)$
    \State $w \gets 1$ if $v \in \{1, \bot\}$ else $0$
    \State $b \gets \textsc{WeakAgreement}(w)$

    \If{$v = 1$}
        \State Return $(1, 1)$
    \ElsIf{$b = 1$}
        \State Return $(1, 0)$
    \ElsIf{$b = 0$}
        \State Return $(0, 1)$
    \Else
        \State Return $(0, 0)$
    \EndIf
\end{algorithmic}
\end{dianabox}

\begin{theorem}
    \textsc{GradedAgreement} satisfies graded agreement.
\end{theorem}

\begin{proof}
    Termination follows from weak agreement's termination. If all honest parties have the same input $1$, then using weak agreement's validity, $v = 1$ and then $b = 1$, so all honest parties return $(1, 1)$. If all honest parties have the same input $0$, then similarly, $v = 0$ and then $b = 0$, so all honest parties return $(0, 1)$. Thus, validity is satisfied.

    We now consider graded agreement. Assume an honest party $p_i$ returns $(1, 1)$. Then $v_i = 1$, so using weak agreement, all honest parties have $v \in \{1, \bot\}$, so they all have $w = 1$, therefore using weak agreement's validity, all honest parties have $b = 1$, and will decide either $(1, 1)$ or $(1, 0)$. We consider the second case where $p_i$ returns $(0, 1)$. This implies that $b_i = 0$. We note that the previous part shows that no honest party could decide $(1, 1)$, so it only remains to show that $(1,0)$ cannot be decided. But if a party $p_j$ decides $(1, 0)$, then $b_j = 1$, which contradicts weak agreement, given that $b_i = 0$. Thus, graded agreement is satisfied.
\end{proof}

\begin{lemma}\label{lemma:graded-comp}
    The protocol is byzantine robust, has round complexity $\roundc_{\gba}(n) = 2\roundc_{\wba}(n)$ and per-party bit complexity $\bitc_{\gba}(n) = 2\bitc_{\wba}(n)$.
\end{lemma}

\subsection{Multivalued Byzantine Agreement}\label{subsec:mvba}

Although our target is binary byzantine agreement, our protocol itself relies on multivalued byzantine agreement, i.e when the value to agree on is an $L$-bit input. We give a simple reduction from multivalued byzantine agreement to binary byzantine agreement, which only uses $\kappa + 1$ calls to binary byzantine agreement.

\begin{dianabox}{\textsc{MVBA}($\vin$)}
\algoHead{Protocol for multi-valued for a party $p_j$ with input $\vin$}
\begin{algorithmic}
    \State Let $h = h_1\ldots h_\kappa$ be the binary representation of the accumulator of $\vin$.
    \InParallel{$i = 1, \ldots, \kappa$}
        \State $d_i \gets \textsc{ByzantineAgreement}(h_i)$
    \EndParallel
    \State $b \gets 1$ if $h = d$ else $0$
    \State $a \gets \textsc{ByzantineAgreement}(b)$
    \If{$a = 0$}
        \State Return $\bot$
    \EndIf
    \If{$h = d$}
        \State Encode $\vin$ into $n$ shares along their witnesses
        \State Send the i-th share to the i-th party
    \EndIf
    \If{$p_j$ receives the valid $j$-th share $s_j$ for the accumulator $h$}
        \State Forward $s_j$ to all parties
    \EndIf
    \If{Received a majority of valid shares for the accumulator $h$}
        \State Decode the shares into $v$
        \State Return $v$
    \Else
        \State Return $\bot$
    \EndIf
\end{algorithmic}
\end{dianabox}

\begin{lemma}
    \textsc{MVBA} satisfies byzantine agreement.
\end{lemma}

\begin{proof}
    Termination follows from the termination of binary byzantine agreement. 

    If all honest parties have the same input $\vin$, the binary agreement's validity ensures that the hash $d$ decided will be the accumulator of $\vin$. So all honest parties will enter the last binary agreement with $b = 1$ and as a consequence, $a = 1$ will be returned. Finally, because all honest parties already have the input matching the accumulator, each honest party will receive a share from each other honest party. Because there is an honest majority, it will be able to reconstruct the input (that it already has) and return it. Thus, validity is satisfied.

    We now consider agreement. If the last binary agreement returns $a = 0$, then using \textsc{BinaryAgreement}'s agreement, all honest parties will return $\bot$ which satisfies agreement. We can now assume that all honest parties have $a = 1$. Because of validity, it implies that at least one honest party has an input whose accumulator is $d$. This honest party will send its share to all other parties, and as a consequence, all honest parties will receive a share from each other honest party. As before, because there is an honest majority, all honest parties will be able to reconstruct the same value and return it. Hence, agreement is satisfied.
\end{proof}

\begin{lemma}\label{lemma:mvba-comp}
    Assuming the binary agreement protocol is byzantine robust,\textsc{MVBA} is byzantine robust, has round complexity $\roundc_{\textsc{MVBA}}(n) = 2 \roundc_{\ba}(n) + 2$ and per-party bit complexity $\bitc_{\textsc{MVBA}}(n, L) = (\kappa+1) \bitc_{\ba}(n) + \mathcal{O}(L + \kappa n \log n)$.
\end{lemma}

\begin{proof}
    Regarding round complexity, the first $\kappa$ calls to \textsc{BinaryAgreement} are in parallel. After the following call, only two more rounds are needed to send and receive the shares, giving the round complexity. 
    Parties run $(\kappa + 1)$ instances of \textsc{BinaryAgreement}. Afterwards, each send at most $2$ shares to each other party. Each share being of size $\mathcal{O}(L/n + \kappa \log n)$, the per-party bit complexity follows.
\end{proof}

\subsection{Group Broadcast}\label{subsec:group-broadcast}

We provide here a simple protocol to implement a group broadcast. A group $G \subseteq \mathcal{P}$ of parties with a byzantine minority want to broadcast the same input. This is done efficiently using errasure codes.

\begin{dianabox}{\textsc{GroupBroadcast}($\vin$, $G$)}
\algoHead{Protocol for group $G$ to broadcast an input $\vin$, run by a party $p$}
\begin{algorithmic}
    \If{$p \in G$}
        \State Let $i$ be the index of $p$ in $G$
        \State Encode $\vin$ into $|G|$ shares and send the $i$-th along with its witness to all parties
    \EndIf
    \If{$p$ receives more than $G/2$ shares with witnesses for the same accumulator}
        \State Decode the shares into $v$
        \State Return $v$
    \Else
        \State Return $\bot$
    \EndIf
\end{algorithmic}
\end{dianabox}

\begin{lemma}\label{lemma:group-broad-prop}
    If $G$ has a byzantine minority and all honest parties in $G$ have the same input, then all honest parties will return this input.
\end{lemma}

\begin{proof}
    If $G$ has an honest majority which has the same input, then all of these parties will send their share for the same accumulator to all parties. As a consequence, all honest parties will receive more than $|G|/2$ shares for the same accumulator, and will be able to decode the input and return it.
\end{proof}

\begin{lemma}\label{lemma:group-broad-comp}\label{lemma:group-broadcast-comp}
    The protocol takes $1$ round, each party in $G$ sends $\mathcal{O}(Ln / |G| + \kappa n \log n)$ bits, regardless of what it receives. Parties outside of $G$ send no message.
\end{lemma}

\begin{proof}
    The protocol ends with each party in $G$ sending their share, and as such takes $n$ rounds. Each party in $G$ sends a share of size $\mathcal{O}(L / |G| + \kappa \log |G|)$ to all $n$ parties. Because $|G| \leq n$, this gives the per-party bit complexity. Parties outside of $G$ do not send any message.
\end{proof}

\subsection{Reliable Group Broadcast}\label{subsec:rel-group-broadcast}

We describe here a protocol similar to the previous one (group broadcast) except that it is reliable: even if the given group $G$ does not have a byzantine minority, or that not all honest parties in $G$ have the same input, all honest parties will return the same output. We note that this protocol can simply be implemented \textsc{GroupBroadcast} followed by \textsc{MVBA}. However, while our use case for \textsc{GroupBroadcast} benefits from it having only parties in $G$ sending a message, our use case for \textsc{ReliableGroupBroadcast} benefits from it being balanced. As such, erasure coding is used slightly differently in this case.

\begin{dianabox}{\textsc{ReliableGroupBroadcast}($\vin$, $G$)}
\algoHead{Protocol for group $G$ to broadcast an input $\vin$, run by a party $p_j$}
\begin{algorithmic}
    \If{$p_j \in G$}
        \State Send the accumulator of $\vin$ to all parties
        \State Encode $\vin$ into $n$ shares and send the $i$-th along with its witness to $p_i$
    \EndIf
    \If{$p_j$ receives the same accumulator $h$ from a majority of $G$ and a valid share $s_j$ with a witness for $h$}
        \State Forward $s_j$ and its witness to all parties
    \EndIf
    \If{$p_j$ receives more than $n/2$ shares and their witnesses for the same accumulator $h$}
        \State Decode the shares into $v$
    \Else
        \State $v \gets \bot$
    \EndIf
    \State Return \textsc{MVBA}($v$)
\end{algorithmic}
\end{dianabox}

\begin{lemma}\label{lemma:reli-group-broad-prop}
    If $G$ has a byzantine minority and all honest parties in $G$ have the same input, then all honest parties will return this input. Additionally, all honest parties will return the same output, regardless of the state of $G$.
\end{lemma}

\begin{proof}
    If $G$ has an honest majority which has the same input, then all of these parties will send their share and the same accumulator to all parties. As a consequence, all honest parties will receive the same accumulator value from a majority of $G$ and a share for this accumulator, which they will forward. Hence, all honest parties will receive shares from all honest parties, which account for more than $n/2$. Thus they will be able to decode the input and \textsc{MVBA}'s validity ensures that it will be returned.

    The last point is a direct consequence of \textsc{MVBA}'s agreement.
\end{proof}

\begin{lemma}\label{lemma:reli-group-broad-comp}
    The protocol has round complexity $\roundc_{\textsc{ReliableGroupBroadcast}}(n) = 2 + \roundc_{\textsc{MVBA}}(n)$ and per-party bit complexity $\bitc_{\textsc{ReliableGroupBroadcast}}(n, L) = \mathcal{O}(L + \kappa n \log n) + \bitc_{\textsc{MVBA}}(n, L)$.
\end{lemma}

\begin{proof}
    Sending and forwarding the shares takes $2$ rounds, followed by the call to \textsc{MVBA}. Each party in $G$ sends a share and witness of size $\mathcal{O}(L/n + \kappa \log n)$ to all $n$ parties, and may forward one such in the second round. Adding the \textsc{MVBA} call, the per-party bit complexity follows.
\end{proof}

\end{document}